\documentclass{ar-1col}
\usepackage[comma]{natbib}
\usepackage{url}
\usepackage{hyperref}
\usepackage{amssymb,amsmath,amsfonts,mathrsfs,bbm,mathabx}
\makeatletter
\def\amsbb{\use@mathgroup \M@U \symAMSb}
\makeatother
\usepackage[bbgreekl]{mathbbol}
\usepackage[english]{babel}
\usepackage{scalerel}

\newcommand{\ve}[1]{\ensuremath{\mbox{\boldmath$#1$}}}
\newcommand{\ma}[1]{\ensuremath{\mathbb{#1}}}
\newcommand{\tr}{{\rm Tr}\,}
\newcommand{\st}{{\rm St}}
\newcommand{\re}{{\rm Re}}

\newcommand{\rep}{{\rm Re_{\rm p}}}{}

\newcommand{\fr}{{\rm Fr}}
\newcommand{\ku}{{\rm Ku}}

\newcommand{\sv}{{\rm Sv}}

\newcommand{\xp}{\ve{x}}
\newcommand{\vp}{\ve{v}}

\newcommand{\etaK}{\eta}
\newcommand{\tauK}{\tau_{\eta}}
\newcommand{\taup}{\tau_{\rm p}}

\newcommand{\ellc}{\ell_0}
\newcommand{\tauc}{\tau_0}
\newcommand{\tauu}{\tau_u}
\newcommand{\minus}{\scalebox{0.75}[1.0]{$-$}}
\usepackage{xcolor}

\newcommand{\bmc}[1]{{{#1}}}

\graphicspath{{./figures/}}
\usepackage{overpic}

\jname{Xxxx. Xxx. Xxx. Xxx.}
\jvol{AA}
\jyear{YYYY}
\doi{10.1146/((please add article doi))}

\begin{document}

% Page header
\markboth{J. Bec, K. Gustavsson, and B. Mehlig}{Statistical models for small particles in turbulence}
% Title
\title{Statistical models for the dynamics of heavy particles in turbulence}

%Authors, affiliations address.
\author{J. Bec$^{1,2}$, K. Gustavsson$^3$, and B. Mehlig$^3$
\affil{$^1$MINES Paris, PSL Research University, CNRS, Cemef, Sophia-Antipolis, France F-06900; email: jeremie.bec@mines-paristech.fr}
\affil{$^2$Universit\'e C\^ote d'Azur, Inria, CNRS, Cemef, Sophia-Antipolis, France, F-06900}
\affil{$^3$Department of Physics, University of Gothenburg, 41296 Gothenburg, Sweden; email: bernhard.mehlig@physics.gu.se}}

%Abstract
\begin{abstract}
When very small particles are suspended in a fluid in motion, they tend to follow the flow.
How such tracer particles are mixed, transported, and dispersed by turbulent flow has been successfully described by statistical models.
Heavy particles, with mass densities larger than that of the carrying fluid, can detach from the flow. This results in preferential sampling, small-scale fractal clustering, and large collision velocities. To describe these effects of particle inertia, it is necessary to consider both particle positions and velocities in phase space.
In recent years, statistical phase-space models have significantly contributed
to our understanding of inertial-particle dynamics in turbulence. These models help to identify the key mechanisms and non-dimensional parameters governing the particle dynamics, and have made qualitative, and in some cases  quantitative predictions.
This article reviews statistical phase-space models for the dynamics of small, yet heavy, spherical particles in turbulence. We evaluate their effectiveness  by comparing their predictions with results from numerical  simulations and laboratory experiments, and summarise their successes and failures.
\end{abstract}

%Keywords, etc.
\maketitle

%% Section 1
\section{INTRODUCTION}
The development of new experimental particle-tracking techniques and 
efficient numerical-simulation methods on parallel computers
have led to a better understanding
of the dynamics, statistics, and geometry of particle transport and mixing
in turbulence~\citep{toschi2009lagrangian}.
When the particles are so small that their inertia is negligible, they follow the 
fluid flow as Lagrangian tracers. Their spatial number density $n(\ve x,t)$ 
obeys the advection equation
\begin{equation}
\label{eq:continuity}
\partial_tn(\ve x,t) + \ve \nabla \cdot \left[\ve u(\ve x,t)\,n(\ve x,t)\right] = 0\,,
\end{equation}
where $\ve u(\ve x,t)$ is the Eulerian fluid velocity at position $\ve x$ and time $t$.
In statistical models of turbulent transport, 
$\ve u(\ve x,t)$ is approximated by a suitably chosen random field
\citep{kraichnan1968small}. 
The 
analysis of such models with tools from statistical physics and dynamical-systems theory has 
provided valuable insight into the anomalous scaling laws governing the transported densities, their geometrical conservation laws \citep{falkovich2001particles}, and their significance for
turbulent mixing \citep{warhaft2000passive,dimotakis2005turbulent} and dispersion \citep{sawford2001turbulent,salazar2009two}.

Many natural problems involve heavier particles\,---\,with mass
densities larger than that of the carrying fluid. Examples are
water droplets in clouds \citep{shaw2003particle,bodenschatz2010can}, 
and dust grains in planet-forming circumstellar 
disks~\citep{birnstiel2016dust}.  Although small in size, such inertial particles can detach from the streamlines of the flow.
Therefore  one must follow the particle velocities $\ve v$, 
in addition \bmc{to} their positions $\ve x$.
Turbulence stretches and folds the patterns formed by particles 
in the phase space spanned by $\xp$ and $\vp$,
so that their velocities become multi-valued in $\xp$-space, or configuration space.
As a consequence, Equation~\ref{eq:continuity} fails to describe the evolution of particle concentrations,
simply because there is no particle-velocity field. Instead, 
the particles form a fractal phase-space attractor, resulting in violent and intermittent fluctuations of particle separations and relative velocities. 
In short, the theoretical
analysis of inertial particles in turbulence requires models that account for the full phase-space dynamics.

Here we review the main insights gained by analysing statistical phase-space models for dilute suspensions of small, yet heavy, spherical particles in homogeneous isotropic turbulence.  We discuss how spatial clustering results from 
centrifugal ejection from vortices and convergence to a fractal attractor. Details depend on the scale at which the patterns are observed, and whether or not the particles settle. 
Folds of fractal phase-space patterns, akin to caustics in geometrical optics,
give rise to anomalous scaling of the particle-velocity structure functions, 
describing large relative velocities between nearby particles which in turn accelerate particle collisions.
Statistical models show that the formation of these caustics depends sensitively on the Stokes number, the particle-inertia parameter. 

The statistical models reviewed in the following are highly idealised. 
We address their strengths and weaknesses by  comparing their predictions
with results obtained by direct numerical simulation of Navier--Stokes turbulence.

For other aspects of the physics and dynamics of dispersed multi-phase flow, we refer the reader
to other reviews. The significance of turbulence-induced droplet collisions
for cloud micro-physics is reviewed by \citet{shaw2003particle} and \citet{grabowski2013growth}.  
\citet{balachandar2010turbulent} discuss the problems arising from two-way coupling between particles and turbulence.
\citet{soldati2009physics} review particles in wall-bounded turbulence, and recent advances in this area  are discussed by \citet{brandt2022particle}, with detailed comparisons between simulation and experimental  results, and between 
point-particle and particle-resolving simulations.
 \citet{fox2012multiphase} reviews 
numerical methods for multiphase flow, and \citet{monchaux2012analyzing} describe experimental advances.  \citet{voth2017anisotropic} review non-spherical particles in turbulence.

%% Section 2
\section{PARTICLE DYNAMICS}
\label{sec:pd}
An incompressible fluid-velocity field $\ve u(\ve x,t)$ solves the Navier--Stokes equations
\begin{equation}
\label{eq:ns}
\ve \nabla \cdot\ve u =0\,, \quad \tfrac{{\rm D}}{{\rm D}t}\ve u \equiv \partial_t \ve u + (\ve u \cdot\ve \nabla)\,\ve u = \tfrac{1}{\varrho_{\rm f}}\,\ve\nabla\cdot\bbsigma\,,
\end{equation}
with fluid density $\varrho_{\rm f}$, and stress tensor $\bbsigma = -p\,\ma 1 + 2\nu{\varrho_{\rm f}} \ma S$. Here $p$ is pressure, $\nu$ is the kinematic viscosity, and
$\ma S=\tfrac{1}{2}(\ma A+\ma A^{\sf T})$ is the symmetric part of the fluid-velocity gradient tensor $\ma A$ with elements $A_{ij} = \partial u_i/\partial x_j$.
Its antisymmetric part is denoted by $\ma{O} = \tfrac{1}{2}(\ma{A}-\ma{A}^{\sf T})$.
The particles impose boundary conditions upon Equation~(\ref{eq:ns}).
For solid particles,
the fluid velocity on any point of the particle surface must equal the particle velocity of this point.

The hydrodynamic force $\ve F_{\rm h}$ on a particle is given by the integral
of the normal component of $\bbsigma$ over the particle 
surface. This non-linear coupling between particle and fluid dynamics poses fundamental difficulties for modeling.
One way out is to rely on particle-resolving numerical simulations 
\citep{tenneti2014particle,maxey2017simulation}.  For turbulent 
suspensions with many particles, this is still very challenging.
An alternative  is to use empirical force models, obtained either by fitting simulation results for a single particle to a model \citep{goossens2019review}, or by solving the Navier--Stokes equations in perturbation theory.
The advantage of empirical parameterisations is that one can go beyond
the perturbative limit. Disadvantages are that results are uncertain outside the fitting range,
and that this procedure 
does not yield immediate insight into the mechanisms at play.

A standard \bmc{perturbative} method is to 
neglect the convective term in Equation~\ref{eq:ns}\bmc{, starting} from the time-dependent Stokes equation. \bmc{I}ncluding the gravitational acceleration $\ve g$, \bmc{one obtains f}or a small sphere of radius $a$ and velocity $\ve v$
\citep{maxey1983equation,landau1987hydrodynamics}:
\begin{eqnarray}
\ve F_{\rm h}&=& \tfrac{4\pi}{3}\varrho_{\rm f}a^3 \tfrac{{\rm D}}{{\rm D}t}\ve u(\xp,t) -6\pi\nu \varrho_{\rm f}a \big[\vp - \ve{u}(\xp,t)\big] -\tfrac{4\pi}{3}\varrho_{\rm f}a^3\,\ve{g} \nonumber\\
&&- \tfrac{2\pi}{3}\varrho_{\rm f}a^3\tfrac{{\rm d}}{{\rm d}t}\big[\vp -\ve u(\xp,t)\big] -6 \sqrt{\pi\nu}\varrho_{\rm f} a^2\scaleobj{0.76}{\int_0^t} \tfrac{{\rm d}s}{\sqrt{t-s}} \tfrac{\rm d}{{\rm d}s} \big[ \vp - \ve{u}(\xp,s) \big] \,.
\label{eq:bbo}
\end{eqnarray}
On the r.h.s.~we have the pressure-gradient force, Stokes' force, Archimedes' force which combines with gravity $\ve F_g=m_{\rm p} \ve g$ to the buoyancy force 
(with particle mass $m_{\rm p}$ and gravitational acceleration $\ve g$), the added-mass force,
and the history force arising from $\partial_t\ve u$ in Equation~\ref{eq:ns}.
Equation~\ref{eq:bbo} is frequently used to model the dynamics of particles in turbulence \citep{brandt2022particle}, sometimes emphasising the significance of the history 
force~\citep{daitche2011memory,olivieri2014effect,guseva2016history,prasath2019accurate}.
However, convective inertia -- neglected in Equation~\ref{eq:bbo} -- 
weakens history effects \citep{lovalenti1993hydrodynamic}.
Fax\'en corrections to Equation~\ref{eq:bbo} have been considered \citep{maxey1983equation}, but for small particles they are of the same order as shear-induced inertia corrections, neglected in Equation~\ref{eq:bbo}.
Moreover, lift forces are neglected, and it
is hard to justify the above form of the added-mass force for turbulent flow \citep{candelier2023second}.
For these reasons, Equation~\ref{eq:bbo} fails to describe the
dynamics of particles in turbulence in general.

For small spherical particles with mass density much larger than
that of the fluid, all terms except Stokes' force
are negligible, fluid inertia does not matter, and the equation of motion 
simplifies to
\begin{equation}
\label{eq:stokes}
\dot \xp = \vp\,,\quad
 \dot  \vp = -\left[ \vp - \ve{u}(\xp,t) \right]/\taup +  \ve g\,.
\end{equation}
Here we included gravity, dots denote time derivatives, and $\taup = m_{\rm p}/ (6\pi \nu \varrho_{\rm f} a)$ is the \textit{Stokes time}.%
\begin{marginnote}[]
\entry{Stokes time}{damping time for the dynamics of inertial particle in the Stokes approximation.}
\end{marginnote}%
It is common to use this sin\-gle\--par\-ti\-cle equation for
many particles, neglecting 
particle-particle interactions (e.g.\ electrostatic or hydrodynamic). 
 Nevertheless, two nearby  particles exhibit correlated motions, because they 
experience correlated fluid velocities.

\begin{textbox}[t]\section{LENGTH AND TIME SCALES OF HOMOGENEOUS ISOTROPIC TURBULENCE}
Turbulent velocity fluctuations are characterised by the root-mean-square velocity $u_{\rm rms} = \langle u_1^2 \rangle^{1/2}$ and the average kinetic-energy dissipation rate $\varepsilon = \nu\,\mathrm{tr}\,\langle \ma{A}^{\sf T} \ma{A} \rangle$.
Turbulent fluctuations span a wide range of length scales, from the
Kolmogorov scale $\etaK = (\nu^{3}/\varepsilon)^{1/4}$  below which viscous dissipation dominates, to the largest scale $L = u_{\rm rms}^3/\varepsilon$ at which the system is driven.
At the spatial scale $\ell$ in the inertial range $\etaK\ll \ell \ll L$, velocity differences grow as $u_\ell \sim (\varepsilon\ell)^{1/3}$ with associated eddy-turnover times $\tau_\ell = \ell/u_\ell \sim \varepsilon^{-1/3}\ell^{2/3}$ \citep{frisch1995turbulence}. Extending this scaling to the edges of the inertial range gives the 
Kolmogorov timescale $\tauK = (\nu/\varepsilon)^{1/2}$ and the large-eddy turnover time $\tau_L = u_{\rm rms}^2/\varepsilon$.  Turbulence intensity is measured by the Taylor-scale Reynolds number $\re_\lambda = u_{\rm rms} \lambda / \nu$, where
$\lambda = u_{\rm rms} / \langle \partial_1u_1^2 \rangle^{1/2}= (15\,\nu/\varepsilon)^{1/2} u_{\rm rms}$ is the Taylor microscale.
\end{textbox}

The non-dimensional parameters of the problem are summarised in Table~\ref{tab:nondimpar}.
As stated above, Equation~\ref{eq:stokes} requires that the particles are small and much heavier than the fluid.
The particle Reynolds number $\rep={v}_{\rm s} a/\nu$ must be small.  Here 
${v}_{\rm s}$ estimates the magnitude of the slip velocity $\ve u-\vp$.
Since the particles detach more easily when $\taup$ is larger, $\rep$ grows as $\taup$ increases.
Also the shear Reynolds number must be small. 
In turbulence, it can be estimated as $\re_s \approx (a/\etaK)^2$.
To neglect molecular diffusion, one must assume a large 
P\'eclet number $\mathrm{Pe}=a^2/(\mathscr{D}\tauK)$
\citep[with diffusion constant 
$\mathscr{D}$; see][]{balkovsky2001intermittent}\bmc{. L}ow mass-loading ($\Phi_{\rm m} =  (\varrho_{\rm p}/\varrho_{\rm f}) n_0 a^3\ll1$ with spatial number density $n_0$) ensures that the particles
do not modify the turbulence \citep{balachandar2010turbulent}.
The Stokes number $\st = \taup/\tauK$
is a measure of particle inertia \citep{snyder1971some}. In the limit $\st\to 0$
at constant non-dimensional settling speed $\sv=g\taup/{u}_\eta$, Equation~\ref{eq:stokes} simplifies to $\ve v = \ve u(\ve x,t)+\sv\, u_\eta\hat{\ve g}$ 
($\hat{\ve g}$~is the direction of gravity). If one keeps the Froude number $\fr = \st/\sv$
constant instead, then $\st\to 0$ constrains the particles to follow the flow, $\ve v = \ve u(\ve x,t)$ (Equation~\ref{eq:continuity}).
Finally, for particles with different sizes, the parameter $\theta{=|\st_1-\st_2|/(\st_1+\st_2)}$ measures Stokes-number differences.

\begin{table}[b]
\caption{\label{tab:nondimpar} Non-dimensional parameters.}
\begin{center}
\begin{tabular}{lll}
\hline
                & $a/\etaK\ll 1$ & small particle size\\
                & $\varrho_{\rm p}/\varrho_{\rm f}\gg 1$ & large mass-density ratio\\
Reynolds number & $\re_\lambda = \sqrt{15}\,\tau_L/\tauK$ & turbulence intensity\\
Particle Reynolds number & $\rep = {v}_{\rm s} a/\nu\ll 1$ &  \raisebox{-2mm}{$\Bigg\}$ convective inertia}\\[-4.mm]
Shear Reynolds number & $\re_{\rm s} =  a^2/(\nu\tauK) = (a/\eta)^2\ll 1$ &                    \\
Mass fraction & $\Phi_{\rm m} =  (\varrho_{\rm p}/\varrho_{\rm f}) n_0 a^3\ll 1 $ & turbulence modification\\
P\'{e}clet number & ${\rm Pe}=a^2/(\mathscr{D}\tauK)\gg 1 $ & effect of diffusion\\
Settling number & $\sv = \taup g /  u_\eta$ &  settling speed \\
Stokes number   & $\st = \taup/\tauK$ &  particle inertia \\
& 
$\theta = |\st_1-\st_2|/(\st_1+\st_2)$  & particle-size dispersion \\
\hline
\end{tabular}
\end{center}
\end{table}
\vfill\eject

Light particles pose additional challenges \citep{mathai2020bubbly}.
Pressure-gradient, added-mass, and buoyancy forces matter.
For bubbles, the boundary conditions at the surface can be different
\citep{legendre1997note}.  The history force is negligible in the high-frequency limit, but lift forces must be accounted for~\citep{mazzitelli2004lagrangian}, and the flow around larger bubbles can become unsteady \citep{mathai2015wake,mathai2016microbubbles}.

%% Section 3
\section{DISSIPATIVE DYNAMICAL SYSTEMS}
\label{sec:dyn_sys}
\subsection{Multifractal phase-space attractor}
\begin{figure}[t]
\noindent\centering\includegraphics[width=13cm,clip]{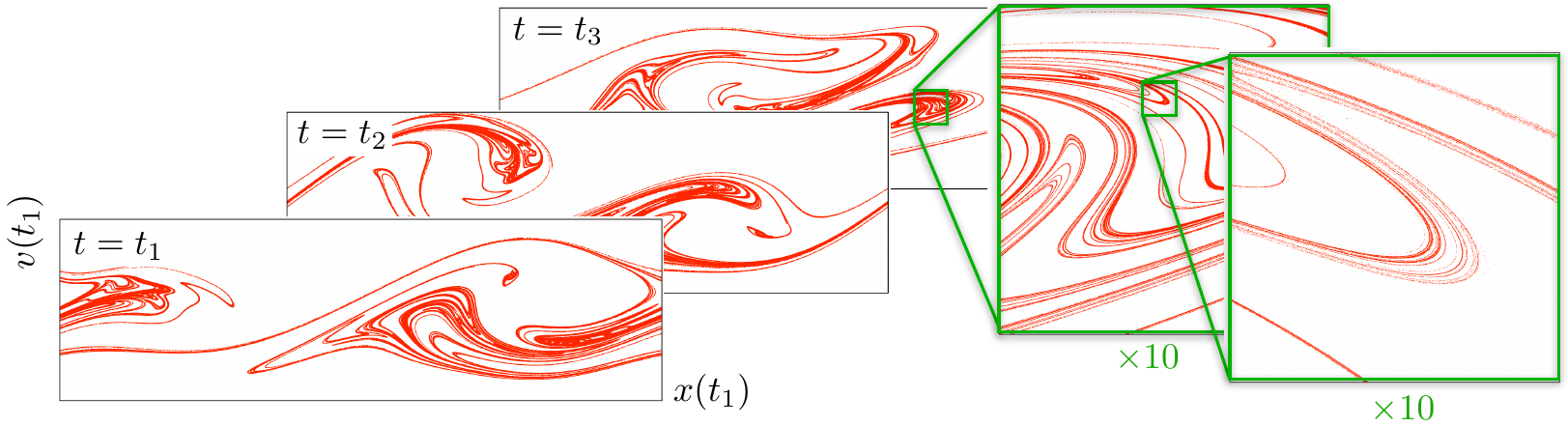}
\vspace{-10pt}
\caption{\textbf{Snapshots of the phase-space attractor} for an idealised model. The equation of motion is a one-dimensional
version of Equation~\ref{eq:stokes}, 
 $\dot x = v$, $\dot v = -[v-u(x,t)]/\taup$, 
with a random velocity field $u(x,t)$ (Section \ref{sec:model}).
Shown are the positions $x(t)$ and velocities $v(t)$ of $10^6$ particles 
at different times (red dots).
The magnifications illustrate fractal clustering.
\label{fig:attractor}}
\end{figure}

Equation~\ref{eq:stokes} defines a dynamical system in  six-dimensional phase space. 
How phase-space structures stretch and contract under the dynamics is measured by the Lyapunov exponents
$\lambda_1\ge \lambda_2 \ge \cdots \ge \lambda_6$. They describe how  nearby particle trajectories diverge
or converge.
The growth rate $\lim_{t\to\infty} t^{-1} \log [\mathscr{V}_{n}(t)/\mathscr{V}_{n}(0)]$ of the volume $\mathscr{V}_{n}(t)$ of an $n$-dimensional phase-space region is $\sigma_n = \lambda_1+\cdots+\lambda_n$
($\sigma_1$ for distances, $\sigma_2$ for areas, \ldots).
 The inertial dynamics of
small particles in incompressible turbulence is chaotic because $\lambda_1>0$. 
At the same time, since $\sigma_6=
 {\nabla_{\!\scriptsize \ve x}\cdot \dot{\ve x} +\nabla_{\!\scriptsize \ve v}\cdot \dot{\ve v}} 
= -{3}/\taup<0$, 
phase-space volumes contract.
In short, the
dynamics of this dissipative chaotic system stretches, folds, and contracts,
causing particle trajectories \bmc{to} 
converge to a phase-space  attractor that evolves in time (\textbf{Figure~\ref{fig:attractor}}).
Dynamical-systems theory allows to analyse this  attractor,
and to draw conclusions about the physical properties of the system (Sections~\ref{sec:preferential}--\ref{sec:relvel}).

Attractors of dissipative chaotic systems are fractal, not smooth.
The distribution of phase-space distances $\delta w =
\sqrt{|\delta \ve x_t|^2 + \tau^2|\delta \ve v_t|^2}\big/\ell$ has power-law form,
Prob$(\delta w < \delta) \sim \delta^{D_2}$ for $\delta \ll 1$, with time and length
scales $\tau$ and $\ell$, and where $\delta \ve x_t$ is the separation between two nearby particles, and $\delta \ve v_t$ their relative velocity. The exponent $D_2$ is the fractal correlation dimension of the phase-space
attractor.  More generally, the attractor is characterised by a spectrum $D_p$ of fractal dimensions
\citep{grassberger1983generalized,hentschel1983infinite}, defined in terms of the fraction of particles $\mathscr{M}_\delta(\ve x_t, \ve v_t,t)$
contained in
a small ball of radius~$\delta$ around 
$(\ve x_t,\ve v_t)$.
For positive integer $p$, the $p$-th moment of  $\mathscr{M}_\delta$ is the probability of finding $p+1$ particles within $\delta$ from each other.
The average  of  $\mathscr{M}^p_\delta(\ve x_t, \ve v_t,t)$ along trajectories
 has power-law form  in the statistically steady state,  $\langle \mathscr{M}^p_\delta\rangle \sim \delta^{\xi_p}$ for $\delta \ll 1$.
The exponents $\xi_p \equiv p\,D_{p+1}$ define the fractal dimensions.
In analogy with multifractal turbulence models, one expects that
$\xi_p$ are non-linear, concave functions of $p$
\citep{paladin1987anomalous,meibohm2020fractal}.
This implies that the phase-space distribution is \textit{intermittent}.
\begin{marginnote}[]
\entry{Intermittent mass distribution}{the
probability density of $\mathscr{M}_\delta$ has non- Gaussian tails.}
\end{marginnote}%
This results in large fluctuations of particle separations and relative velocities (Sections  \ref{sec:fractal} and \ref{sec:relvel}).

\begin{textbox}[t]%
\section{MEASURING FRACTAL DIMENSIONS OF SPATIAL PATTERNS\label{sec:gR}}
The radial distribution function $g(R)$ measures the probability of finding two particles at distance $R$ in configuration space. In three spatial dimensions, 
the function is defined as $g(R)= {(4\pi R^2})^{-1} \tfrac{{\rm d}}{{\rm d}R} \langle \hat{\mathscr{M}}_R\rangle$, and it is normalised such that
$g(R)=1$ for uniformly distributed particles.
When the particles form fractal clusters $g(R)$ grows as a power law $\propto R^{\hat{D}_2-{3}}$ at small distances, with spatial correlation dimension $\hat{D}_2<3$.
 A fractal dimension introduced by
 \citet{kaplan1979chaotic}, $\hat{D}_{KY}$, measures the dimensionality of spatial regions that neither expand nor contract. This dimension
can be calculated from the spatial Lyapunov exponents.
Under quite general circumstances one can show that $\hat{D}_{KY}=\hat D_1$ \citep{ledrappier1988dimension}.
\end{textbox}

\subsection{Multi-valued velocities and projection}
To determine the spatial distribution of particles, one must project the
phase-space patterns to configuration space ($\ve x$-space).
 A fold of a smooth phase-space curve causes a divergence of the projected spatial density, 
a catastrophe or caustic just as in geometrical optics \citep{berry1980catastrophe}, and in the mass distribution in the early universe \citep{zeldovich1970gravitational}.
The phase-space patterns of heavy particles in turbulence are fractal. Nevertheless
their folds \citep[fractal catastrophes, see][]{meibohm2020fractal}
result in large spatial particle-number densities \citep{crisanti1992lagrangian,martin1994accumulation,wilkinson2005caustics}.  

Between two caustics, the projection to configuration space is many-to-one \citep{falkovich2002acceleration}. 
\begin{marginnote}[]%
\entry{Caustic}{fold location of a smooth phase-space curve in configuration space.}
\end{marginnote}%
If the particles approach on different branches of the phase-space attractor
 (\textbf{Figure~\ref{fig:attractor}}), 
they can collide at high velocities, a process called sling effect by \citet{falkovich2002acceleration}.
\begin{marginnote}[]%
\entry{Sling effect}{mechanism by which particles collide at high velocities, facilitated by caustics.}
\end{marginnote}%
In  \textbf{Figure~\ref{fig:attractor}},
fold caustics are points on the $x$-axis.
In reality, fold caustics  are surfaces in configuration space around regions of multi-valued particle velocities.
Higher-order catastrophes can occur,  such as cusp catastrophes, swallow-tails, and umbilics
\citep{berry1980catastrophe}.
As the attractor folds, caustics occur at rate $\mathscr{J}$, increasing
the number of phase-space branches with different velocities. This is balanced by the contraction of the dynamics in the velocity direction,
so that a steady state emerges.

Spatial distributions of particles in turbulence inherit their fractal properties  from the fractal phase-space attractor.
The fraction of particles $\hat{\mathscr{M}}_\delta(\ve x_t,t)$ in a sphere of non-dimensional radius $\delta$ around $\ve x_t$
obeys
$\langle \hat{\mathscr{M}}_\delta^{p}\rangle \sim \delta^{{p} \hat{D}_{{p}+1}}$ for $\delta \ll 1$, with spatial fractal dimension $\hat{D}_{{p}+1}$.
The first moment defines the the spatial correlation dimension ${\hat D}_2$ (Section \ref{sec:fractal}).
For a typical projection of a generic attractor to a three-dimensional subspace, the projected correlation dimension obeys $\hat{D}_2 = \mbox{min}\{D_2,{3}\}$ 
\citep{hunt1997how,bec2008stochastic}. \citet{meibohm2020fractal} showed that 
this saturation of $\hat{D}_2$ at the spatial dimension is due to caustics,
they result in a uniform particle distribution at large $\st$. 
Projection formulae for other fractal dimensions are not known.
Settling changes fractal clustering in configuration space 
(Section~\ref{sec:fractal}), the spatial patterns become anisotropic 
\citep{bec2014gravity,gustavsson2014clustering,ireland2016effect_b}.

Particles with different Stokes numbers cluster on different fractal attractors, shifted with respect to each other \citep[][Section~\ref{sec:fractal}]{chun2005clustering,bec2005clustering}.
The phase-space attractors are shifted too,
affecting the relative particle-velocity distribution (Section~\ref{sec:relvel}).

Spatial clustering has been analysed using Voronoi tessellations of thin slices of particle positions \citep{monchaux2010preferential}.
This provides a robust measure of particle clustering. But it does not yield information about the fractal dimensions.

%% Section 4
\section{STATISTICAL MODELS}
\label{sec:model}
Statistical models avoid some of the challenges posed by \textit{ab-initio simulations} of particles in turbulence. The idea is to study the solutions of Equation~\ref{eq:stokes} with a prescribed fluid velocity $\ve u(\ve x,t)$, using mathematical analysis, statistical closure, or Monte-Carlo simulation.  
The goal is twofold. First, there is no other way 
at the moment of studying how the 10$^{8}$ water droplets in a cubic metre of a turbulent atmospheric cloud determine its radiative properties \citep{schneider2017climate}. Second, the analysis of idealised statistical models can explain 
the key mechanisms and 
non-dimensional parameters that determine the particle dynamics.
\begin{marginnote}[]
\entry{Lagrangian fluid elements}{infinitesimal fluid elements that follow the flow like tracers.}
\end{marginnote}

This approach has been successful for inertialess tracer particles.
Their turbulent mixing is understood using Lagrangian models, which approximate the turbulent
velocity fluctuations seen by \textit{Lagrangian fluid elements}~\citep{pope1994lagrangian}. A different point of view was taken by \citet{kraichnan1968small}. He analysed how
the spatial concentration $n(\ve x,t)$  of tracer particles develops according
to Equation~\ref{eq:continuity}, using a  statistical model for the Eulerian field  $\ve u(\ve x,t)$. 
The theoretical analysis of Kraichnan's model allows to relate the scalar concentration fluctuations to the relative motion of the tracers~\citep{falkovich2001particles}.
\begin{marginnote}[]
\entry{Kraichnan model}{approximates $\ve u(\ve x,t)$ as a Gaussian random 
field, which is white-noise in time with power-law correlations in space.}
\end{marginnote}

Heavy particles that do not follow the flow call for different levels of sophistication in the modelling. One can solve the Navier--Stokes equations without the particles by direct numerical simulation and integrate Equation~(\ref{eq:stokes}) with the numerically computed~$\ve u(\ve x,t)$ \citep{riley1974diffusion,squires1991preferential,sundaram1997collision}. This 
 method is called \textit{one-way-coupled} DNS \citep{elgobashi2019direct}.  
Although not {\em ab-initio}, it has significantly advanced 
our understanding of the problem~\citep{brandt2022particle}.

\begin{textbox}[b]\section{MATCHING STATISTICAL MODELS TO TURBULENCE} 
In \textbf{KS-models}, a discrete  set of Fourier modes $k_n$ is chosen to approximate the turbulent kinetic-energy spectrum $E(\ve k)\simeq  1.5\, \varepsilon^{-2/3} |\ve k|^{-5/3}$ (upper and lower cutoffs are matched to $\eta$, and to the large scale $L$). Time correlations are imposed by the 
frequency spectrum $\omega_n =\omega_0 \sqrt{k_n^3 E(k_n)}$, where $\omega_0$ determines the Kubo number \citep{vosskuhle2015collision}.  The Stokes number is not defined in terms of $\tauK$, but using a typical wave number \citep{ijzermans2010segregation}, introducing  an unknown factor in the definition of $\st$. The three parameters $u_0$, $\ellc$, $\tauc$ of the \textbf{single-scale model} 
are matched as follows. The magnitude of 
actual turbulent fluid-velocity gradients is $\sim\!\tauK^{-1}$, and Lagrangian fluid elements separate and decorrelate on time scales $\sim\!\tauK$.  The 
single-scale model exhibits the same behaviour for $\tauc \gg \tauK$.  In this large-$\ku$ limit, the Eulerian time scale $\tauc$ does not matter, and one 
matches the model time-scale $\ellc/u_0$ to the Kolmogorov time $\tau_\eta$ 
of the turbulent flow.  To model single-particle statistics (Section~\ref{sec:preferential}), the remaining parameter is chosen as $u_0=\sqrt{3}u_{\rm rms}$, or equivalently $\ellc=\lambda$, the Taylor microscale.  For the dynamics of particle separations (Sections \ref{sec:fractal} and \ref{sec:relvel}), one matches $\ellc$ to the length scale where the smooth dissipation range turns into the inertial range, of the order $\ellc\sim 10\etaK$.  For \textbf{KE-models}, one matches the eddy-turnover time $\tau_\ell$ to turbulence, as well as the correlation times $\tau_O$ and $\tau_S$ of vorticity and strain.  The resulting model contains unknown higher-order structure functions of fluid-velocity differences. They are approximated in terms of two-point functions.  The latter are taken from DNS~\citep{bragg2014new_a,bragg2014new_b}.
\end{textbox}
An alternative to DNS is to develop and analyse idealised models for the turbulent flucutations.
\citet{maxey1986gravitational} used an ensemble of stationary cellular flows.
\cite{saffman1956collision}  averaged over Gaussian-distributed fluid-velocity gradients $\ma A$ of time-independent linear flows $\ve u(\ve x) = \ma A\, \ve x$.
\citet{falkovich2007inertial} used a one-dimensional model where $A = \partial u/\partial x$ follows a telegraph process. Another possibility is to  represent $\ve u(\ve x,t)$ as a sequence of non-linear velocity fields $\ve u_t(\ve x)$ changing at random times $t$ \citep{pergolizzi2012etude}.  
The most common approach, however, is to approximate the turbulent velocity by an incompressible random field. To ensure incompressibility, it is convenient to represent the fluid velocity as the curl of a random vector potential, $\ve u(\ve x,t)=\ve\nabla\wedge\ve \Psi(\ve x,t)$.  The vector potential $\ve \Psi(\ve x,t)$  is constructed from random spatial Fourier modes. %
\begin{marginnote}[]%
\entry{KS model}{approximates $\ve u(\ve x,t)$ as a Gaussian random function
with power-law correlations in space and time.}
\end{marginnote}%
Kinematic-simulation (KS) models \citep{fung1992kinematic,ducasse2009inertial,ijzermans2010segregation} approximate turbulent inertial-range scaling by imposing a power-law dependence of the correlation function $\langle \Psi_i(\ve x,t)\Psi_j(\ve x',t')\rangle$ upon $\ve x-\ve x'$, as in the Kraichnan model. While Kraichnan's model is white noise in time, KS models have a non-zero decay time.

To describe the particle dynamics in the dissipative range of turbulence, 
one can simplify the model further, keeping only a single spatial scale $\ellc$.
Time correlations can be generated by Ornstein-Uhlenbeck processes, resulting in
components $\Psi_j$ of the vector field that are Gaussian distributed with zero mean and correlations~\citep{maxey1987gravitational,pinsky1995model,sigurgeirsson2002model,bec2005clustering,bec2008stochastic,gustavsson2016statistical}, %
\begin{marginnote}[]%
\entry{Single-scale model}{approximates $\ve u(\ve x,t)$ as a Gaussian random function
with the single spatial scale $\ellc$.}
\end{marginnote}%
\begin{align}
\label{eq:Acorr} \left\langle \Psi_i(\ve x,t)\Psi_j(\ve x',t')\right\rangle
=\mathscr{C}\,(u_0\,\ellc)^2\,\delta_{ij}\, \exp\big[-{|\ve x-\ve x'|^2}/(2\ellc^2)-{|t-t'|}/{\tauc}\big]\,.
\end{align}
The normalisation constant $\mathscr{C}$ is chosen so that $\langle |\ve u|^2\rangle=u_0^2$.
The model has two time scales: the Eulerian correlation time $\tauc$ 
and the Kolmogorov time 
$\tauK=(\tr\,\langle \ma{A}^{\sf T} \ma{A} \rangle)^{-1/2}$
which evaluates to $\tauK=\ellc/(u_0\sqrt{{5}})$.  The ratio of these time scales defines the 
Kubo number ${\rm Ku} =\frac{u_0\tauc}{\ellc}\propto \frac{\tauc}{\tauK}$~\citep{wilkinson2007unmixing}.
In order to match the model to turbulence, one considers the 
limit of large Kubo numbers, where the dynamics becomes independent of 
$\tauc$. The remaining
parameters, $u_0$ and $\ellc$, can then be matched to the relevant scales of the turbulent flow.  
\bmc{In} this case, any dependence on $\re_\lambda$ is contained in the choice of $u_0$ and $\ellc$.  
The model allows mathematical analysis in the limit of small but finite $\ku$
 using perturbation theory \citep{gustavsson2016statistical}.
The white-noise limit (the limit  $\ku\to 0$ at constant $\ku\, \st$)
  can be analysed
using diffusion approximations \citep{wilkinson2007unmixing}.
The persistent limit of $\ku\to \infty$
at constant $\st$ can in some cases be treated analytically~\citep{derevyanko2007lagrangian,pergolizzi2012etude,meibohm2023caustics}.

The single-scale  model assumes that the particle dynamics is mainly determined by the smooth
part of the turbulent velocity correlation function.  At large Stokes numbers or settling velocities, the inertial range
 may influence the particle dynamics, and this can lead to new parameter dependencies, for example on $\re_\lambda$.
KS models do have a range of spatial scales, but they do not account for the sweeping of small turbulent eddies by large ones. Instead,  large eddies drag Lagrangian fluid elements through smaller eddies.
As a consequence, fluid-velocity gradients decorrelate too quickly compared with turbulence \citep{vosskuhle2015collision}.
Both models have  Gaussian fluid-velocity fluctuations. This means that
they  neglect violent fluctuations of the fluid-velocity gradients corresponding to persistent regions of anomalously high turbulent vorticity. Under which circumstances this matters  is discussed in the following Sections.

Other models for dispersed multiphase flow are related to Langevin models for the Lagrangian dynamics of fluid elements~\citep{pope1994lagrangian}. Approximate Langevin equations for inertial particles \citep{minier2016statistical} allow to compute the distribution of $\ve x(t)$, $\ve v(t)$, and $\ve u(\ve x(t),t)$ by Monte-Carlo simulation. The models reproduce single-time, single-particle statistics, but  they are not designed to describe two-point statistics such as relative particle-velocities. Multi-valued particle velocities can be represented in a heuristic fashion, by decomposing the particle velocities into two contributions, a smooth velocity field plus spatially uncorrelated fluctuations intended to represent the effect of caustics \citep{fevrier2005partitioning}.
Alternatively, one may consider models for the distribution of $\ve x(t)$
and $\ve v(t)$ alone. One can derive a kinetic equation (KE) for $P(\ve x,\ve v;t)$ if one assumes that the turbulent velocity fluctuations are Gaussian. Keeping only the leading terms in a moment expansion, one obtains:
\begin{equation}
\left\{\partial_t + v_i \partial_{x_i} 
- \tfrac{1}{\taup} \partial_{v_i} v_i
- \tfrac{1}{\taup}\partial_{v_i} \left[\partial_{v_j} \mu_{ij}(\ve x,t) + \partial_{x_j} \lambda_{ij}(\ve x,t)-\kappa_i(\ve x,t)\right]\right\}P(\ve x,\ve v;t)=0\,.
\end{equation}
Here $\mu_{ij}$, $\lambda_{ij}$, and $\kappa_i$ 
are unknown coefficients that require \textit{closure} approximations, 
and summation over repeated indices is implied.
This approach yields accurate numerical descriptions of single-particle properties, also in spatially inhomogeneous turbulent flow \citep{reeks2021development}. \cite{zaichik2003pair} derived analogous equations for two-particle statistics. 
Such \textit{KE models} can be closed using DNS results for the unknown 
\begin{marginnote}[]%
\entry{Closure}{approximate evaluation of unknown terms in a moment expansion.}
\end{marginnote}%
\begin{marginnote}[]
\entry{KE model}{kinetic equation for distribution of particle positions and velocities, requiring closure.}
\end{marginnote}%
coefficients~\citep{bragg2014new_a,bragg2014new_b}, 
to obtain numerical results for the statistics of separations and relative velocities between heavy particles in turbulence.
In Sections~\ref{sec:fractal} and~\ref{sec:relvel}, we compare the corresponding results to those of statistical models based on Equation \ref{eq:Acorr}. 
In the white-noise limit, the kinetic equation for particle separations and relative velocities
reduces to a Fokker-Planck equation, of similar form as the one-dimensional equation  analysed by \cite{gustavsson2008variable}.

%% Section 5
\section{PREFERENTIAL SAMPLING }
\label{sec:preferential}
\subsection{Maxey's centrifuge}
\label{sec:pref}
\begin{figure}[t]
\begin{overpic}[width=16cm]{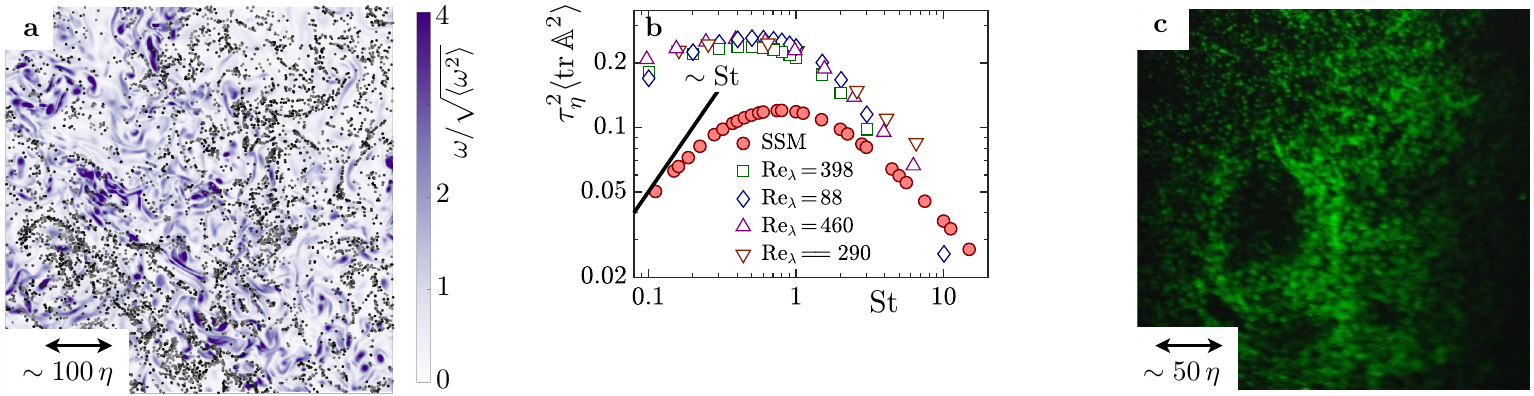}
\end{overpic}
\caption{\textbf{Preferential sampling}. 
{\bf a}~Heavy particles avoid vortices. Shown are DNS results for particle positions
in a thin slice of three-dimensional configuration space. Different levels of gray reflect particle
depth within the slice [data from \cite{bec2014gravity} with $\st=1$ and $\re_\lambda\approx460$]. 
Also shown is the magnitude $\omega$ of vorticity (see text).  
{\bf b}~Average of $\tr\ma A^2$ along inertial-particle paths versus $\st$.
DNS results from \citet{ireland2016effect_a} ($\scriptstyle \Box$, $\Diamond$), \citet{bec2014gravity} ($\scriptstyle\bigtriangleup$, $\scriptstyle\bigtriangledown$ ),  data from simulations of the single-scale model (Equation~\ref{eq:Acorr}) for $\ku=10$ (\raisebox{-1.pt}{\large $\bullet$}), and the prediction of perturbation theory $\tauK^2\langle \tr\ma A^2\rangle \propto \st$ (solid line). 
 {\bf c}~Voids in the spatial distribution of cloud droplets. 
Data from \citep{karpinska2019turbulence}. Image obtained from E. Bodenschatz.
\label{fig:preferential}}
\end{figure}

Heavy particles can detach from the flow,
they are not Lagrangian fluid elements.
This leads to a bias in the statistical properties of the fluid-velocity gradients evaluated along particle paths \citep{maxey1987gravitational}, called 
\textit{preferential sampling}.  Maxey
obtained an approximate equation-of-motion by expanding
Equation~\ref{eq:stokes} at small $\st$. When $\sv=0$, this yields $\tfrac{{\rm d}}{{\rm d}t} \ve x \approx  \ve v_{\rm p}(\ve x,t) =  \ve u -\taup \tfrac{\rm D}{{\rm D}t} \ve u$.
Since the effective particle-velocity field $\ve v_{\rm p}(\ve x,t)$ is compressible,
\begin{marginnote}[]
\entry{Maxey's centrifuge}{particle inertia expels heavy particles from
vortices.}
\end{marginnote}
$\ve \nabla \cdot \ve v_{\rm p} = -\taup\tr \ma A^2 = -\taup
\big(\tr\ma S^2-\tr\ma O^{\sf T}\ma O \big) \neq 0$,
he concluded that small-$\st$ particles are more likely to explore
 the sinks of $\ve v_{\rm p}(\ve x,t)$,
regions of large strain and small vorticity. This \textit{centrifuge} effect 
is illustrated in \textbf{Figure~\ref{fig:preferential}a}, 
showing a snapshot of particle positions together with the magnitude of vorticity $\omega \equiv \ve |\nabla \wedge \ve u(\ve x,t)| = \sqrt{2\tr\ma O^{\sf T}\ma O}$. 
Heavy particles tend to avoid connected regions of high vorticity 
of linear sizes smaller than $\sim{10\eta}$ (Section~\ref{sec:model}). 

\citet{falkovich2004intermittent} predicted that $\langle \tr \ma A^2\rangle {\tauK^2}\propto \st$ along particle paths. This is consistent with a perturbative analysis 
of the single-scale model neglecting multi-valued particle velocities, which gives
$\langle \tr \ma A^2\rangle\tauK^2{\propto}\st$  
for small $\ku$ and $\st$  \citep{gustavsson2016statistical}.
\textbf{Figure~\ref{fig:preferential}b} shows $\langle \tr \ma A^2\rangle$ from DNS and simulations of the single-scale model.
The results are qualitatively similar, but there is one important  difference: the model
underestimates preferential sampling \citep{gustavsson2016statistical} because turbulence has more persistent regions of much larger vorticity that act as efficient centrifuges 
(\textbf{Figure \ref{fig:preferential}a}).
Note that the DNS results decrease more slowly than predicted
by perturbation theory as $\st\to 0$. 
 The reason for this difference is not understood.
\textbf{Figure~\ref{fig:preferential}b} also shows that the centrifuge effect becomes weaker at larger $\st$.
This is expected, because the  history of the fluid-velocity gradients seen by the particle
 matters more than the gradients at the present position
\citep{gustavsson2008variable,bragg2015relationship}.
Note also that the DNS results in \textbf{Figure~\ref{fig:preferential}b} show a weak 
dependence on $\re_\lambda$. A possible explanation is that at larger $\re_\lambda$
particles are more likely to be expelled from straining regions \citep{ireland2016effect_a}.

An alternative preferential-sampling mechanism was suggested by \citet{coleman2009unified}: at small $\st$, heavy particles cluster near points of zero fluid acceleration, where $\tfrac{{\rm D}}{{\rm D}t}\ve u=0$. At small
Stokes numbers, this is consistent with Maxey's small-$\st$ expansion. 
At large $\st$, by contrast, this expansion fails, and preferential sampling of low-acceleration regions becomes negligible \citep{bragg2015mechanisms}.

Quantifying preferential sampling in laboratory experiments is hard because one must simultaneously track particles and fluid-velocity gradients.
The measurements of \citet{gibert2012where} and \citet{petersen2019experimental} show a certain bias towards straining regions at intermediate Stokes numbers.

\subsection{Settling}
Preferential sampling has a significant effect on the gravitational settling of particles in turbulence. Averaging
Equation~\ref{eq:stokes} leads to $\langle \ve v\cdot \hat{\ve g}\rangle = 
\taup\,g+\langle \ve u\cdot \hat{\ve g}\rangle$,
where $\taup\, g$ is the Stokes settling speed in a still fluid.
\citet{maxey1987gravitational} suggested that turbulence enhances settling, because particles expelled from vortices preferentially sample down-welling regions where $\ve u\cdot \hat{\ve g}>0$. This 
\textit{preferential sweeping} is confirmed by DNS \citep{wang1993settling,bec2014gravity,rosa2016settling}, as well as  experiments \citep{aliseda2002effect,good2014settling,petersen2019experimental}.\begin{marginnote}[]
\entry{Preferential sweeping}{particles preferentially sample down-welling regions of turbulent flow.}
\end{marginnote}
The relative increase in settling speed 
is largest for settling numbers $\sv\approx 1$, when the settling velocity is of the order of the smallest turbulent eddies.
For large $\sv$, 
\citet{good2014settling} observed that turbulence can have the opposite
effect, reducing the settling speed.
What causes this reduction is not understood \citep{rosa2016settling},
but model calculations using cellular flow fields 
confirm this behaviour ~\citep{delillo2008sedimentation,martins2008terminal}.

\subsection{Large-scale preferential sampling}
\label{sec:large_scales}
The mechanisms reviewed in Section~\ref{sec:pref}
explain preferential sampling in the dissipative range of turbulence,
leading to spatial patterns on scales up to $\ellc\sim 10\eta$. 
\textbf{Figure~\ref{fig:preferential}a} shows
strong preferential sampling also at much larger scales. 
Similar structures -- termed \textit{cloud holes} by
\cite{karpinska2019turbulence} -- are observed in the spatial patterns of cloud 
droplets (\textbf{Figure~\ref{fig:preferential}c}). 
It is plausible that these voids are caused by rare fluctuations
of turbulent vorticity, intense and persistent vortex tubes that
manage to expel particles with moderate inertia.

At large Stokes numbers, inertial-range statistics may become important. The mean-squared particle velocity, for example, is of the order of $\langle v_1^2\rangle \sim u_{\rm rms}^2 /(1+\taup/\tauu)$,
where $\tauu \!\sim\! \tau_L \!=\! \tauK\re_\lambda/\sqrt{15}$ is the Lagrangian correlation time of the 
large-scale 
fluid velocity \citep{abrahamson1975collision,ireland2016effect_a}.
In other words,  $\langle v_1^2\rangle$ obtains a ${\rm Re}_\lambda$ dependence due 
to large-scale motion. This effect can be accounted 
for by setting $u_0=\sqrt{{3}}\,u_{\rm rms}$ and ${\ellc} \!\sim\! u_{\rm rms}\,\tau_L$ in the single-scale model, Equation~\ref{eq:Acorr}.
A related choice of scales is discussed in Section~\ref{sec:fractal}, intended to describe inertial-range 
clustering in terms of a scale-dependent Stokes number  $\st_L \sim \taup/\tau_L$. The only $\re_\lambda$-dependence is contained in the choice of the time scale $\tau_L$.

The settling velocity of heavy particles, on the other hand, exhibits an intricate $\re_\lambda$-dependence~\citep{bec2014gravity}, that cannot be described by simply changing a time scale.  The reason is preferential sampling:
rapidly settling particles explore vertical fluid-velocity correlations at spatial separations much larger than the dissipation scale.
The settling particles experience an effective horizontal flow that is compressible, and collect in the sinks of this flow.
A Kraichnan model for the effect of large eddies
on the settling speed predicts $\langle \vp\cdot \hat{\ve g} \rangle/v_{\rm s}-1 \propto \re_\lambda^{3/4}\fr^{5/2}\st^{-2}$, in good agreement with DNS \citep{bec2014gravity}.

%% Section 6
\section{FRACTAL CLUSTERING}
\label{sec:fractal}
\subsection{Multiplicative amplification}
\label{sec:mechanisms}

Tracer particles 
in a compressible flow 
form fractal spatial patterns. 
The sum of the spatial Lyapunov exponents is negative,
so that the Kaplan-Yorke dimension $\hat D_{\rm KY}$ is smaller than 
the spatial dimension \citep{sommerer1993particles,falkovich2001particles,toschi2009lagrangian}.
% see Section~\ref{sec:dyn_sys}.
In incompressible flow, tracer particles cannot cluster because the spatial Lyapunov exponents 
sum to zero. The
infinitesimal volume
$\hat{\mathscr{V}}_3=\mbox{det}[\ve R_1,{\ve R_2},\ve R_3]$ spanned by 
three separation vectors $\ve R_\alpha$ between four nearby particles  
neither contracts nor expands.
% (Section~\ref{sec:dyn_sys}).

Inertial particles cluster even in incompressible flow. 
To understand the mechanism, consider
how $\hat{\mathscr{V}}_3$ changes.  If the particles remain close to each other, one can show that $\hat{\mathscr{V}}_3(t+\delta t) = \det[\ma I+\ma Z(t)] \hat{\mathscr{V}}_3(t)$. Here, $\ma Z$ is the matrix of particle-velocity gradients $Z_{ij} = \partial v_i/\partial x_j$.  Since the product of many random factors $\det[\ma I+\ma Z(t)]$
tends to be smaller than unity, spatial volumes contract. This amplifies local particle-number density fluctuations, resulting    in spatial clustering 
by \textit{multiplicative amplification} \citep{gustavsson2016statistical}.

The dynamical equation for $\ma Z$ follows
from Equation~\ref{eq:stokes} by differentiation,
$\tfrac{\rm d}{{\rm d}t} \ma Z =  (\ma A-\ma Z)/\taup -\ma Z^2$ 
\citep{falkovich2002acceleration}. It must be integrated alongside the particle motion, therefore fractal clustering is determined by the history of the fluid-velocity gradients experienced by the particles \citep{gustavsson2008variable,bragg2015relationship}.  
In the white-noise limit (Section~\ref{sec:model}), fractal clustering is entirely due to multiplicative amplification \citep{wilkinson2007unmixing}.

\subsection{Correlation dimension}
\label{sec:smallscale}
The spatial correlation dimension $\hat D_2$ has
a straightforward interpretation in terms of the pair-correlation function, and quantifies
the effect of fractal clustering on the collision rate (Section~\ref{sec:relvel}). Unfortunately, however, $\hat D_2$ is hard to calculate.
Systematic perturbation expansions for the spatial correlation dimension
are known only for special cases, such as the white-noise limit.
To make matters worse, the perturbation series diverge asymptotically.
In addition, it appears that the perturbation theory
misses non-analytical contributions proportional to the 
rate of caustic formation~\citep{gustavsson2015analysis}.
\begin{figure}[t]
\begin{overpic}[width=15cm]{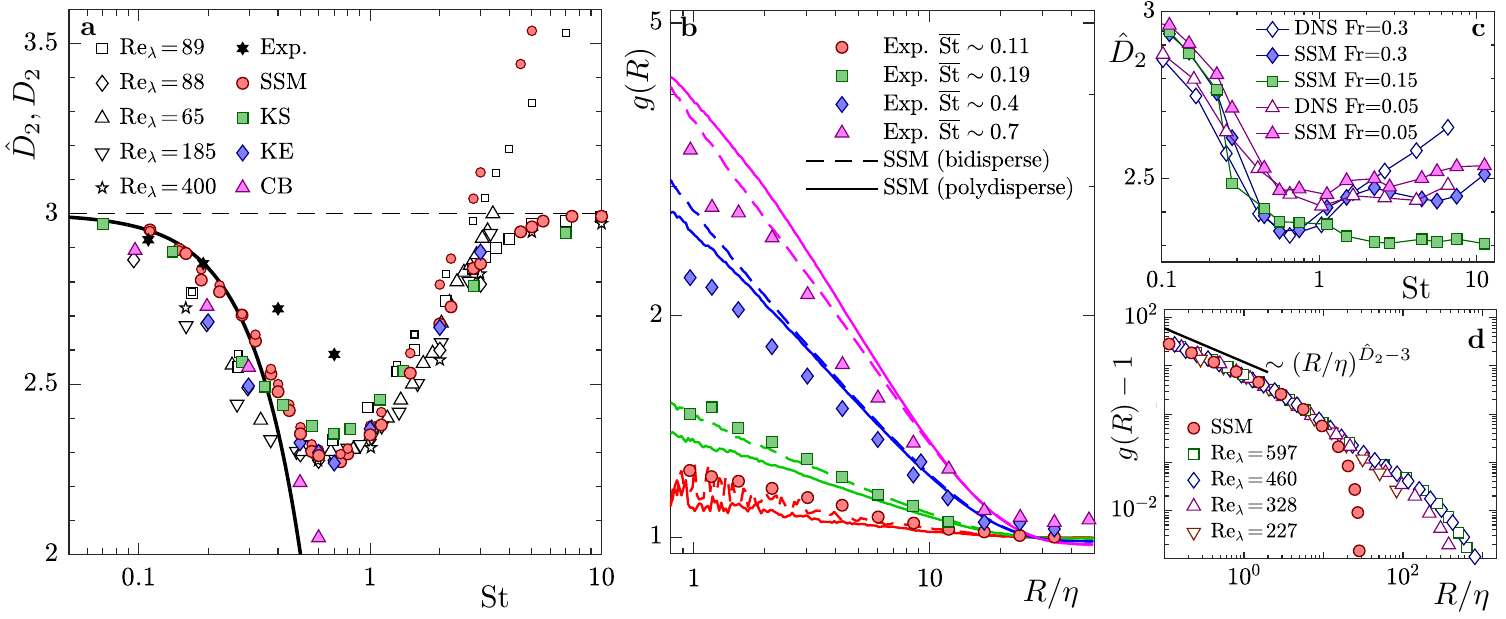}
\end{overpic}
\caption{\label{fig:clustering}
 \textbf{Fractal clustering}.
{\bf a} Correlation dimension for identical particles for  $\sv=0$.
Spatial $\hat{D}_2$ (large symbols), phase-space $D_2$ (small symbols) versus $\st$.
\textbf{DNS results}:
${\rm Re}_\lambda=89$ \citep{bhatnagar2018statistics},
${\rm Re}_\lambda=88$ \citep{bragg2014new_a},
${\rm Re}_\lambda=65,185$ \citep{bec2007heavy}, 
${\rm Re}_\lambda=400$ \citep{bec2010intermittency}.
\textbf{Model predictions}: KE-model, evaluated by \cite{bragg2014new_a} using DNS for the unclosed terms (blue,$\blacklozenge$), simulations of the single-scale model (SSM, Equation~\ref{eq:Acorr}) for $\ku=10$ (red, $\bullet$ and \raisebox{-1.5pt}{{\Large $\bullet$}}), together with a fit to the small-$\st$ asymptote (solid line).  Also shown are simulations of the KS-model~\citep{ijzermans2010segregation} (green, 
\textcolor{black}{$\blacksquare$}),  see text. Small-$\st$ theory of \citet{chun2005clustering} evaluated by \citet{bragg2014new_a} using DNS for the unclosed terms (magenta, \textcolor{black}{$\blacktriangle$}).  \textbf{Experimental data} extracted from panel~{\bf b} by fitting the data in the range  $1\le R/\eta\le 10$ (\raisebox{-1.5pt}{\Large $\star$}).
{\bf b}~Pair correlation functions $g(R)$ for polydisperse 
suspensions of droplets in air turbulence.  Laboratory experiments \citep{saw2012spatial_b}, symbols, compared with
predictions of single-scale model (SSM) $\ku=10$ and $\ell_0/\eta=20$.  First: 
for particles of two different sizes (dashed lines) with $\overline{\st}\sim 0.11, \theta\sim 0.91$; $\overline{\st}\sim 0.19, \theta=0.53$; $\overline{\st}\sim 0.4, \theta=0.37$; $\overline{\st}\sim0.7, \theta=0.43$,  read off from Fig.~7 in \citet{saw2012spatial_b}. 
The average Stokes number is defined as $\overline{\st}=\overline{\tau}_{\rm p}/\tauK$,
where $\overline{\tau}_{\rm p}$ is the harmonic mean of the different Stokes times, see \citep{meibohm2017relative}.
Second: 
distribution of particle sizes extracted from their Fig.~3 (solid lines).
{\bf c} Effect of settling, $\hat{D}_2$ for different values of $\fr$, DNS results from \citet{bec2014gravity}, single-scale model (SSM) for $\ku=10$ and $\ellc/\eta=10$.  {\bf d}~Inertial-range clustering, pair-correlation function $g(R)$ versus $R$ for $\st=1$. DNS at $\re_\lambda=597$ \citep{ireland2016effect_a}, ${\rm Re}_\lambda=460$ \citep{bec2014gravity}, ${\rm Re}_\lambda=328$ \citep{ariki2018scale}, and ${\rm Re}_\lambda=227$ \citep{ray2011preferential}. Single-scale model for $\ku=10$ and $\ellc/\eta=20$, and small-$R$ asymptote of $g(R)$, solid line~(Section~\ref{sec:dyn_sys}).}
\end{figure}

\textbf{Figure \ref{fig:clustering}a} shows DNS measurements of 
$\hat{D}_2$ for different $\re_\lambda$. Also shown are results
of model calculations: numerical evaluations 
for the KE-model carried out by \cite{bragg2014new_a} using DNS for the unclosed terms, for the single-scale model (Equation~\ref{eq:Acorr}), and for the KS model~\citep{ijzermans2010segregation}. 
As pointed out in Section~\ref{sec:model}, the definition of the Stokes number for the KS-model contains an unknown factor.
Here we simply rescaled $\st$ for the KS-model data (by a factor of $1.4$) to fit the location of the minimum in the DNS data in \textbf{Figure \ref{fig:clustering}a}. 
Overall, we observe good agreement between  model and  DNS results.
In particular, $\hat D_2$ depends only weakly on $\re_\lambda$.
This is expected, because $\hat D_2$
characterises the small-$R$ asymptote of $g(R)$, dominated  by
particle pairs that have spent a long time together 
at small relative velocities, but only the tails of the relative-velocity distribution are sensitive to $\re_\lambda$.
Therefore the small-$R$ behaviour 
of $g(R)$ is insensitive to $\re_\lambda$-corrections.
\textbf{Figure \ref{fig:clustering}a} demonstrates that fractal clustering is strongest at $\st\approx 1$,
as expected,
 because turbulent fluctuations affect the particles most when $\taup$ and $\tauK$ 
are comparable.

The largest deviations between model and DNS results are observed
at small Stokes numbers. When $\st\ll1$, 
one may replace the inertial-particle dynamics by advection in an effective velocity field $\ve v_{\rm p}(\ve x,t)$ with $\st$-dependent compressibility 
(Section~\ref{sec:preferential}).  This suggests that $3-\hat{D}_2\propto \st^2$ \citep{balkovsky2001intermittent,elperin2002clustering}.  
\cite{chun2005clustering} came to the same conclusion, expanding the relative-particle dynamics for small $\st$. There is, however, no agreement on the numerical prefactor. 
An improved version of the small-$\st$ theory by \cite{chun2005clustering}
is also shown in \textbf{Figure \ref{fig:clustering}a}, evaluated by \citep{bragg2014new_a} using
DNS for the unclosed terms. The result is consistent with a $\st^2$-dependence, as predicted by perturbative analysis of the single-scale model \citep{gustavsson2016statistical}.
But the DNS results are not. The reason for this discrepancy is unclear. 

A large-deviation argument indicates that $ 3-\hat D_2 = 2(3-\hat{D}_{\rm KY})$ at small $\st$
\citep{fouxon2011construction}. The same result holds in the white-noise limit  \citep{gustavsson2015analysis,gustavsson2016statistical}.  DNS results are consistent with this prediction \citep{bec2006lyapunov,bec2007heavy,calzavarini2008dimensionality}.  

Regarding the phase\--space correlation dimension $D_2$, 
we observe good overall agreement between DNS and model simulations 
(\textbf{Figure \ref{fig:clustering}a}), although 
small differences at large $\st$ remain unexplained.
We also see that $\hat{D}_2=\min\{D_2,3\}$, consistent with the projection formula discussed in Section \ref{sec:dyn_sys}. %
\begin{marginnote}[]
\entry{Caustics and $\hat{D}_2$}{caustics explain that the spatial correla- tion dimension $\hat{D}_2$ equals the spatial di- mension for~$\st\!>\!\st_{\rm c}$.}
\end{marginnote}

\textbf{Figure \ref{fig:clustering}b} shows the experimental data of \citet{saw2012spatial_b} for $g(R)$, averaged over particle sizes, and
results of simulations of the single-scale model, in good agreement with experiments.
Averaging over different particle sizes reduces the local slope
${\rm d}\log g(R)/{\rm d}\log R$, reducing clustering. This could explain why experimental data (with size dispersion) display a larger value of $\hat D_2$ than observed in DNS (\textbf{Figure \ref{fig:clustering}a}). 
\citet{larsen2018fine-scale} observed spatial clustering of cloud droplets in the dissipation range. Their results for $g(R)$ are broadly consistent 
with the model predictions: below $R\sim\ellc$, $g(R)$ increases as $R$ decreases.
However, since  the Stokes numbers are very small ($\st\approx 10^{-2}$), the clustering is so  weak that its effect on the collision rate (Section~\ref{sec:relvel}) is negligible. %
\begin{marginnote}[]%
\entry{Size dispersion}{bidisperse sus- pensions contain particles
of two different sizes (with different Stokes numbers, $\st_1$ and $\st_2$).
Polydisperse suspensions involve a distribution
of many different sizes.}
\end{marginnote}%

Settling affects fractal clustering, because settling  changes how particles sample fluid-velocity gradients. Since gravity breaks the rotational symmetry, clustering becomes anisotropic \citep{bec2014gravity,gustavsson2014clustering}, so that the pair correlation depends on the angle $\varphi$ between $\ve g$ and $\ve{\hat{R}}=\ve R/R$.  This affects only the prefactor, not the exponent: $g(R) = C(\varphi) R^{\hat{D}_2-{3}}$ \citep{bhatnagar2020statistics}.
\textbf{Figure \ref{fig:clustering}c} shows $\hat{D}_2$ versus $\st$ for different values of the settling number,
for DNS and model simulations. Overall we observe good agreement.
At small $\st$, fractal clustering decreases slightly as the effect of gravity increases
 (smaller $\fr$). But at large $\st$,
settling increases clustering by multiplicative amplification -- rapidly settling
particles experience the flow as white noise \citep{bec2014gravity,gustavsson2014clustering}.
Note that the correlation dimension depends non-monotonically on $\fr$ in this regime.

\citet{lu2010clustering_a} measured 
the effect of electrical charges on spatial clustering of droplets in turbulence.  They determined $g(R)$ for highly charged droplets of equal parity, and found that the resulting electrostatic interactions affect the pair correlation function $g(R)$  at length scales below $R/\etaK \sim 4$, for $\st=0.3$ and $\re_\lambda=80$. At present there is no systematic study of how the spatial clustering depends on the amount of \bmc{charge}, on $\st$, and on $\re_\lambda$.

\subsection{Inertial-range clustering}
\label{sec:irc}
The fractal dimensions $\hat{D}_{\rm KY}$ and $\hat{D}_2$ discussed above measure small-scale fractal clustering in the dissipative range of turbulence.
In turbulent flow, spatial clustering is observed not only in the dissipative range but also
in the inertial range. However, as seen from \textbf{Figure \ref{fig:clustering}d}, the radial distribution function $g(R)$ no longer behaves as a power-law when $R\gg\eta$. In the inertial range, the turnover time $\tau_{{R}} = ({R}^2/\varepsilon)^{1/3}$ depends on the spatial scale ${R}$. 
%(\textbf{Figure~\ref{fig:preferential}a}). 
Based on this observation, 
\citet{falkovich2003statistics} suggested 
that the radial distribution function behaves as $\log g(R)\propto \st_R^2 \sim \taup^2/R^{4/3}$ \citep{balkovsky2001intermittent}. 
The scale-dependent Stokes number
arises naturally in white-noise models. Simulations of the Kraichnan model
yield $\log g(R)/\log R \simeq \hat{D}_2(\st_R)-3 \propto \st_R^2$ \citep{bec2008stochastic}.
To which extent time correlations change this picture is unclear. KS models 
cannot answer this question,  because they do not account for sweeping.
In particular, the degree of large-scale clustering 
obtained in KS-model simulations depends on the parameter $\omega_0$,
that is on the Kubo number (Section~\ref{sec:model}), 
see \cite{chen2006turbulent}.
There is numerical evidence
for the scaling  $\log g(R)\propto R^{-4/3}$ from DNS \citep{bragg2015mechanisms,ariki2018scale}. KE-models yield the same inertial-range scaling, but they do not capture the Stokes-number dependence~\citep{bragg2015mechanisms}.

An alternative picture was proposed by \citet{bec2007heavy}. They argued and observed in DNS that $g(R){R^2}$ depends on a non-dimensional ejection rate $\Gamma(R,\taup) \propto u_{\rm rms}\varepsilon^{1/3}\taup/R^{5/3}$. How to reconcile this behaviour with the 
above $\st_R$-scaling remains an open question. The function $\Gamma(R,\taup)$ prescribes the scale-dependence not only of $g(R){R^2} = \tfrac{\rm d}{{\rm d}R}\langle \hat{\mathscr{M}}_R\rangle$, but of the whole probability distribution $p(\hat{\mathscr{M}}_R)$. In DNS, deviations from a uniform distribution are significant in the two tails of this distribution. The form of the tails is explained by the statistical ejection model of \citet{bec2007toward}. They found that $p(\hat{\mathscr{M}}_R) \propto \hat{\mathscr{M}}_R^\alpha$ at small $\hat{\mathscr{M}}_R$,  and $\propto \exp(-c\,\hat{\mathscr{M}}_R \log \hat{\mathscr{M}}_R)$ for large $\hat{\mathscr{M}}_R$, where $\alpha$ and  $c$ depend only on $\Gamma(R,\taup)$. 

A certain form of scale invariance is recovered in the distribution of large void sizes.
DNS of two-dimensional  \citep{boffetta2004large,goto2006self} and three-dimensional
turbulence  \citep{yoshimoto2007self}, as well as experimental data \citep{sumbekova2017preferential} suggest that the probability distribution of very large void sizes behaves as  a power-law.
However, different authors obtain different power-law exponents.

%% Section 7
\section{CAUSTICS AND RELATIVE VELOCITIES}
\label{sec:relvel}
\begin{figure}
\begin{overpic}[width=15cm]{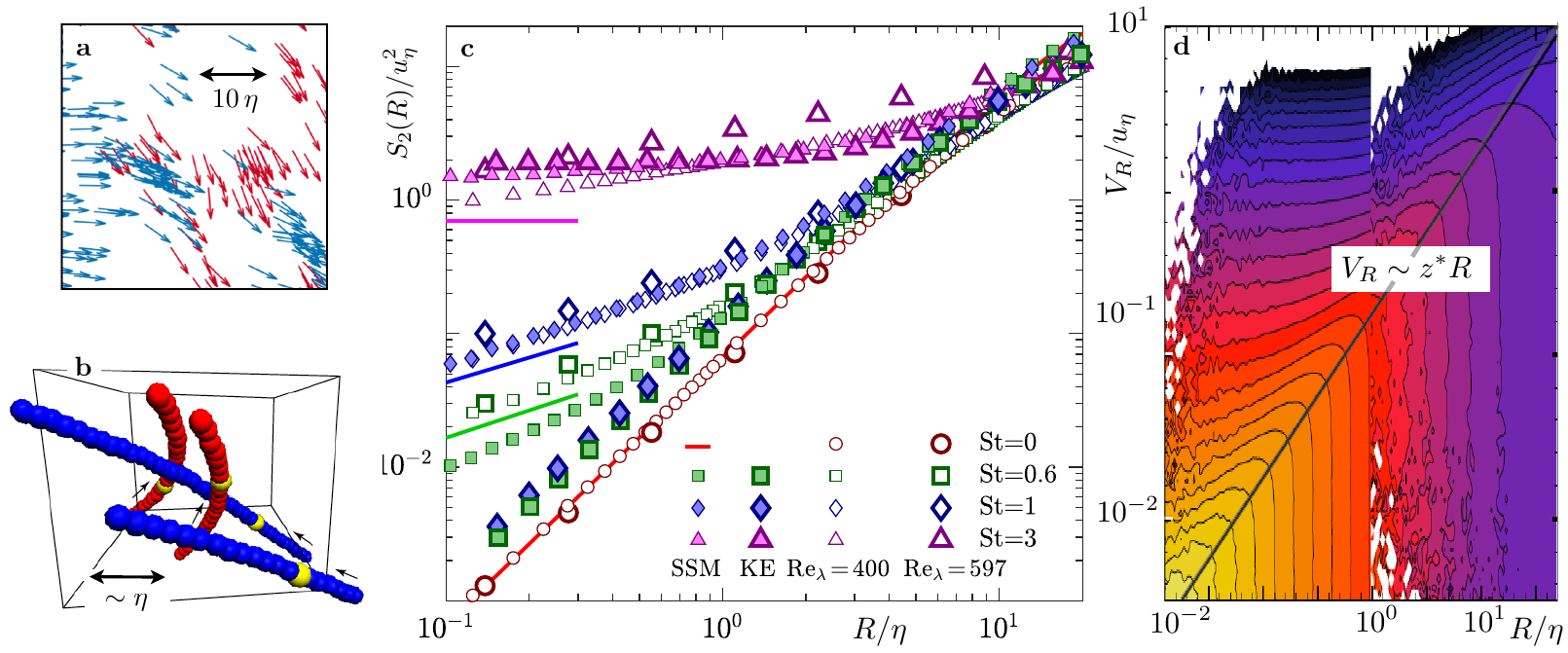}
\end{overpic}
\caption{\label{fig:rel_vel} \textbf{Relative velocities}.
{\bf a}~Multi-valued particle velocities (red and blue arrows) from a thin slice of a three-dimensional DNS \citep{bec2014gravity}.
The two colours distinguish two particle clouds with different histories.
{\bf b}~Experimental observation of two nearby particle pairs (red and blue) with anomalously large relative velocity, adapted from Fig.~2 in \cite{bewley2013observation} with permission.
{\bf c}~Particle-velocity structure function $S_2(R)$
versus $R$ for $\sv=0$.
DNS from \cite{bec2010intermittency} [$\re_\lambda=400$, small empty symbols], and \cite{bragg2014new_b} [$\re_\lambda=597$, large empty symbols],
single-scale model (SSM, Equation \ref{eq:Acorr}) with $\ku=10$ and $\ell_0/\eta=15$
(small filled symbols),  $\st\!=\!0$-result from Equation \ref{eq:Acorr} (solid red line),
small-$R$ scaling prediction~\ref{eq:moments} (lines),
and KE-model predictions  evaluated by \cite{bragg2014new_b}
(large filled symbols)  
{\bf c}~Joint distribution of $R$ and $V_R$ from DNS, 
adapted from Fig.~3(b) in \citep{bhatnagar2018statistics}. 
The matching line $V_R \sim z^\ast R$ (see text) 
is shown as a thick solid line. }
\end{figure}
Particle inertia causes large fluctuations of relative particle velocities $\ve V=\ve v_2-\ve v_1$
in turbulence. The reason is that phase-space manifolds fold at caustic singularities, resulting in multi-valued particle 
velocities (Section~\ref{sec:dyn_sys}).
This is illustrated in \textbf{Figure~\ref{fig:rel_vel}a}, which shows  inertial-particle velocities in a thin slice of configuration space.
The two colours distinguish two particle clouds that move on different branches of the phase-space attractor. \textbf{Figure~\ref{fig:rel_vel}b} shows measured droplet trajectories  in turbulent air \citep{bewley2013observation}. As in panel {\bf a}, the two colours distinguish particle pairs 
originating from different regions in phase space, allowing the
pairs to  pass close to each other at anomalously large
relative velocities.

\subsection{Rate of caustic formation}
\label{sec:J}
The rate of caustic formation $\!\mathscr{J}\!$ depends sensitively on $\st$,
because dimensional analysis shows that the required strain for a caustic to form is 
proportional to $\st^{-1}$. At small $\st$, the  rate of caustic formation
 is therefore determined by
the tails of the turbulent velocity-gradient distribution \citep{meibohm2023non,baetge2023quantitative}.
In the white-noise limit, diffusion approximations for the Gaussian single-scale model yield
$\minus\log (\mathscr{J}\!\taup) \propto (\ku\,\st)^{-1}$ for $\ku\,\st\ll 1$  \citep{wilkinson2005caustics}.
In the persistent limit of the Gaussian model, the rate has a different form,
$\minus\log(\mathscr{J}\!\taup) \propto \st^{-2}$ \citep{falkovich2002acceleration,derevyanko2007lagrangian}.
Kinematic simulations by \cite{ducasse2009inertial}
and DNS by \cite{bhatnagar2022rate}
yield $\minus\log(\mathscr{J}\!\taup) \propto \st^{-1}$. In general,
one expects $\minus\log(\mathscr{J}\!\taup) \propto {F}(\st)$ where ${F}\to\infty$ as $\st\to 0$.
The form of ${F}$
depends on the tails of the velocity-gradient distribution. 
The key point is nevertheless that caustics form at small $\st$, leading to multivalued
particle velocities. For very small $\st$, however, phase-space
contraction reduces the effect of caustics, when $\mathscr{J}\!\taup\ll 1$.
\citet{vosskuhle2013prevalence} concluded that caustics have a significant effect on the
collision rate for $\st {\gtrsim} 0.3$. 
 Exceptions to this rule are model flows
with bounded fluid-velocity gradients, 
where $\!\mathscr{J}\!$ vanishes below a certain Stokes number~\citep{ijzermans2010segregation}.

\begin{marginnote}[]
\entry{Caustic formation}{Fold caustics
form when the par\-ti\-cle- ve\-lo\-ci\-ty gradients diverge, equivalently when the
infini- tesimal volume $\hat{\mathscr{V}}_{3}$ collapses.}
\end{marginnote}
\citet{perrin2014preferred} found that caustics 
tend to occur in regions
of large turbulent strain.  \citet{meibohm2021paths,meibohm2023caustics} explained this by    
computing the optimal path to caustic formation
in the persistent limit,
at weak particle inertia, and $\sv=0$. 

\subsection{Relative particle velocities}
\label{sec:rvm}
\citet{volk1980collisions} estimated the relative velocities between identical particles as $\langle V_R^2\rangle \sim\varepsilon\taup$.  
Related expressions for particles with different sizes were suggested by
\citet{mizuno1988grain}. These estimates 
fail to describe the relative particle-velocities 
at small spatial separations, because the derivation
ignores the dissipation-range dynamics \citep{pan2010relative}.
DNS results indicate that the relative-velocity moments $m_n(R) \equiv \int_{-\infty}^\infty {\rm d}V_R |V_R|^p P(R,V_R)$ depend sensitively on both $R$ and $\st$. \textbf{Figure~\ref{fig:rel_vel}c}
shows DNS results for the relative particle-velocity structure function 
$S_2(R) \equiv m_2(R)/m_0(R)$. Also shown are results of model calculations, for
the KE model as evaluated by \citep{bragg2014new_b},
and for the single-scale model based on Equation~\ref{eq:Acorr}.
For advected particles ($\st=0$), $S_2(R) \propto R^2$  in the smooth dissipative range.
Particle inertia enhances the relative particle velocities, more so
at smaller separations. The single-scale model describes this behaviour well, although
the weak dependence on $\re_\lambda$ seen for DNS is not captured by the model. The KE-model works well at $\st=3$, but fails at smaller values of the Stokes number. \citet{bragg2014new_b} 
suggest that this may be due to Gaussian decoupling  approximations used in evaluating the model (Section~\ref{sec:model}).

The structure function exhibits anomalous scaling at small $R$, $S_{n}(R)\sim R^{\zeta_n}$.
\citet{gustavsson2014relative} derived an approximate  theory 
for the exponents $\zeta_n$
by matching the tails of the joint distribution $P(R,V_R)$ at $|V_R| = z^\ast R$, with matching
scale $z^\ast\propto \taup^{-1}$ (\textbf{Figure~\ref{fig:rel_vel}d}).  
Recognising that $P(R,V_R)$  must have algebraic large-$V_R$ tails at small $R$, determined by the phase-space correlation dimension $D_2$ (Section \ref{sec:dyn_sys}), one finds
\begin{align}
\label{eq:moments}
m_n(R) 
{\ \simeq\ }  b_n(\st) (R/\etaK)^{p+D_2-1} + c_n(\st) (R/\etaK)^{{2}}
\end{align}
for small separations $R$.  
Equation~\ref{eq:moments} predicts that
$m_0(R)$ is proportional to $ R^{\mbox{\scriptsize  min}\{D_2,{3}\}-1}$ 
\bmc{for small enough $R$.} 
The scaling of the radial
distribution function $g(R)= m_0(R)/R^{2}{\propto} R^{\hat{D}_2-{3}}$ implies that $\hat{D}_2=\mbox{min}\{D_2,{3}\}$ (Sections~\ref{sec:dyn_sys} and \ref{sec:fractal}). 
Using this scaling of $m_0(R)$, 
Equation~\ref{eq:moments} yields a \textit{bifractal} law  for the 
scaling exponents of the structure functions $S_n(R)$, namely
$\zeta_n = \mbox{min}\{n,3-\bmc{\hat D}_2{(\st)}\}$ for $n>0$, 
consistent with the DNS results for $n=2$ shown in \textbf{Figure~\ref{fig:rel_vel}c}.
Note that the saturation of $\zeta_n$ at  large~$n$ is a consequence
of caustic singularities. Analogous behaviour is observed in Burgers turbulence, where the singularities are shocks.
Note also that \citet{simonin2006connection} suggested that $S_2(R)$ approaches a constant as $R\to 0$,
reflecting random uncorrelated motion (Section~\ref{sec:model}).
For $n>0$, this is true only when
 $\st$ is so large that 
$D_2>3$.
\begin{marginnote}[]
\entry{Bifractal scaling}{The scaling exponents of structure functions equal  $\zeta_n \!=\! h_1 n$ for $n\!<\!n_\star$, and $\zeta_n \!=\! h_2 n \!+\! (h_1\!-\!h_2)n_\star$ for $n\!>\!n_\star$. Scaling laws of this kind arise in the $\beta$-\-model for turbulent intermittency and in Burgers turbulence
\citep{frisch1995turbulence}.  For the bifractal law corresponding to Equation~\ref{eq:moments}, $h_2=0.$
}
\end{marginnote}

The two terms on the r.h.s.\ of Equation~\ref{eq:moments} correspond to different histories of the relative dynamics.  The first term is a smooth contribution, it comes from particles that stayed together in the past because they approached along the same phase-space branch.
The second is a singular contribution, corresponding to particles that followed different phase-space branches 
(\textbf{Figure~\ref{fig:attractor}}). Its $R$-dependence, 
$\propto R^{2}$, reflects the uniform distribution
of nearby  particles that came from far apart with 
uncorrelated initial conditions.

There is no consensus regarding the particle relative-velocity distribution. It has been modeled as 
 lognormal \citep{carballido2010relative}, or with stretched
exponential tails \citep{pan2014turbulence_b}.
The matching ansatz outlined above yields 
$P(R,V_R) R^{{2}-D_2} = f(V_R/R)$ for small distances,
with  $f(y) \approx {\rm min}\{y,y^{D_2-{4}}\}$.
DNS results \citep{perrin2015relative,bhatnagar2018statistics} at small 
$\re_\lambda$ support this prediction.
%  (\textbf{Figure~\ref{fig:rel_vel}d}).
\citet{saw2014extreme} measured $P(R,V_R)$ for water droplets in air turbulence. Their results do not
show algebraic tails, likely because  $R/\eta$ \bmc{is} too large.
Measurements of \citet{hammond2021particle} 
\bmc{with hollow glass spheres show} much broader distributions than those
of \citet{saw2014extreme}, although their experiments explored similar parameter ranges.
When the Stokes number is large enough, the analysis of \cite{gustavsson2008variable} suggests
that relative velocities are determined by the history of inertial-range fluctuations,
resulting in
$P(R,V_R) \propto \exp\{-[V_R^2/(\taup\varepsilon)]^{2/3}\}$.
This is consistent with the results of
\citet{pan2010relative,pan2013turbulence}.
% At very large $\st$, when $\taup\gg\tau_L$, the relative-velocity distribution becomes Gaussian, and $\langle V_R^2\rangle \propto \st$ \citep{pan2010relative,pan2013turbulence,gustavsson2014relative}.

\citet{bhatnagar2020statistics} concluded that settling does not change the relative dynamics
very much, for identical particles.
In poly-disperse suspensions, by contrast, settling is expected to have a stronger effect on the relative velocities.

For $\sv=0$, 
the effect of size dispersion upon the distribution of $V_R$, 
is discussed by \cite{pan2014turbulence_a}. 
In essence, particles with different Stokes numbers tend to have larger relative velocities.
\citet{meibohm2017relative} show that
 $S_n(R)$ approaches a constant for
$R<R_\theta \propto \theta$ with 
$\theta{=|\st_1-\st_2|/(\st_1+\st_2)}$,
causing an enhancement of the collision rate (Section~\ref{sec:cr}).

Electrostatic charges affect the relative particle velocities \citep{alipchenkov2004clustering,lu2015charged}, but it is not yet clear how electrostatic interactions modify the relative-velocity theory reviewed above.

\subsection{Models for the collision rate}
\label{sec:cr}
\citet{saffman1956collision} computed how turbulence accelerates collisions between small cloud droplets.
They assumed that the droplets follow the flow ($\st=0$) and that they
are uniformly distributed. 
Linearising the fluid velocity,
assuming time-independent gradients, \citet{saffman1956collision}
averaged over the gradients to obtain
$\mathscr{R}_{\rm ST} = C_1 n_0 (a_1+a_2)^3 /\tauK$.
Comparing with the collision rate for droplets settling in a quiescent fluid,
$\mathscr{R}_{g}\simeq n_0 v_s (a_1+a_2)^2$ with settling speed $v_s = \tfrac{2\varrho_{\rm p}}{9\varrho_{\rm f}{\nu}}|a_{2}^2-a_{1}^2|g$, leads to 
two conclusions: $\mathscr{R}_{\rm ST}$ dominates for similar-sized droplets, and it increases proportional to the turbulent shear rate.

To determine the effect of particle inertia,
\citet{sundaram1997collision} integrated Equation~\ref{eq:stokes} for fluid velocities obtained from turbulent DNS. They found
that the  collision rate
\begin{align}
\label{eq:RSt}
\mathscr{R}({\st})&=-n_0\lim_{R\to a_1+a_2}\int_{-\infty}^0\!{\rm d}V_R \,V_R\, P(R,V_R)
\end{align}
with number density $n_0,$ depends sensitively on the Stokes number. For identical particles this is explained by the fact that caustics allow for large relative velocities, and that the rate of 
caustic formation depends sensitively on $\st$.
A model for the $\st$-dependence follows from Equation~\ref{eq:moments},
using $\mathscr{R}({\st}) \approx \tfrac{1}{2} m_1(2a)$. This gives $\mathscr{R}({\st}) \approx \mathscr{R}_{{\rm smooth}} + \mathscr{J}(\st) \mathscr{R}_{\rm kin}$ \citep{falkovich2002acceleration}. Equation \ref{eq:moments}
implies that $\mathscr{R}_{{\rm smooth}} \propto a^{\hat D_2}$,  and $\mathscr{R}_{\rm kin}\propto a^2$, the collision rate in the kinetic limit
\citep{abrahamson1975collision}.
The second term dominates
for small particles and large enough~$\st$.
This occurs already for $\st>0.3$
\citep{vosskuhle2011estimating,vosskuhle2013prevalence}.
\begin{marginnote}[]
\entry{Hydrodynamic interactions}{for two nearby particles,
the presence of the second particle changes the fluid velocity
seen by the first one, causing them to interact.}
\end{marginnote}
\begin{marginnote}[]
\entry{Collision efficiency}{accounts for the effect of
hydrodynamic interactions on the collision rate.}
\end{marginnote}

Some authors write $\mathscr{R}({\st})\!=\!\tfrac{1}{2} g(2a) \langle |V_R| \rangle_{R=2a}$ \citep{hammond2021particle},
in terms of the average $\langle \cdot \rangle_{R=2a}$ 
conditioned on the particle separation at contact.
This does not mean that spatial clustering and  relative velocities contribute in a multiplicative fashion, because  $\langle |V_R|\rangle_{R=2a}$ equals $m_1(2a)/m_0(2a)=m_1(2a)/[(2a)^2g(2a)]$, so  the factor $g(2a)$ 
cancels out, see Fig.~9(d)--(f) in \citep{nair2022effect}.
Equation~\ref{eq:moments} shows
that the collision rate for identical particles
depends on these two effects additively.
Particle-size differences increase the collision rate between small particles because they increase their relative velocity at separations smaller than $R_\theta$, where $S_2(R)$ approaches a constant
(Section \ref{sec:rvm}).
 \citet{pan2014turbulence_c} discuss in detail how size dispersion competes with particle inertia in determining~$\mathscr{R}$. 

The models described above neglect hydrodynamic interactions,
which make it harder  for 
the particles to approach (and to separate).
In order to determine whether the particles collide or not, one must also account for the fact that hydrodynamics breaks down when
the distance between the particle surfaces is
smaller than the mean-free path of the fluid \citep{sundararajakumar1996noncontinuum}. 
For water droplets in air, this changes the interplay between 
$\mathscr{R}({\st})$ and $\mathscr{R}_g$,
and reduce the overall collision rate \citep{dhanasekaran2021collision}.

Hydrodynamic interactions can be incorporated in independent-particle
models in \bmc{a} heuristic fashion, by writing the collision rate
as $e_{\rm c}\, \mathscr{R}({\st})$ \citep{pinsky1999collisions,pinsky2007collisions,devenish2012droplet,pumir2016collision},
where $e_{\rm c}$ is a collision efficiency intended to account for hydrodynamic interactions.
\citet{klett1973theoretical} realised that the collision efficiency of
similar-sized settling droplets depends strongly on $\re_p$, simply because identical droplets do 
not even approach each other for $\rep=0$. 
\citet{klett1973theoretical}  and later studies \citep{pinsky2007collisions,wang2008turbulent}
accounted for hydrodynamic interactions by perturbation theory in $a/R$. This describes the droplet dynamics at separations
of the order of several droplet radii \citep{magnusson2021collisions}, but it does not allow to compute the collision efficiency reliably. 
Studies resolving close droplet approaches were performed only for $\st=0$~\citep{dhanasekaran2021collision,dubey2022bifurcations}.
In summary, it is not understood how the collision efficiency depends on $\re_p$, $\st$, and $\sv$.

Recent experiments \citep{yavuz2018extreme,bragg2021hydrodynamic} measuring spatial clustering of small particles in turbulence 
show that the pair correlation function $g(R)$ is large at very small separations,
much larger
than predicted by the theories summarised in Section \ref{sec:fractal}.
The mechanisms for this strong clustering at very small scales remain to be understood.

%% Summary points
% Summary Points
\begin{summary}[SUMMARY POINTS]
\begin{enumerate}
\item Statistical models provide a qualitative understanding of the
phase-space dynamics of particles in turbulence,
allowing to single out aspects of the 
dynamics that are sensitive to 
details of the turbulent flow. In certain cases, statistical models can be analysed systematically using dynamical-system\bmc{s} and perturbation theories, 
making it possible to 
gain valuable insight into key mechanisms and parameters.
\item Statistical models explain
that dissipative-range spatial clustering of particles in turbulence is determined by the history of
the flow encountered by the particles. 
 To predict small-scale clustering in a quantitative fashion, 
one must therefore take into account that this history is  biased by preferential sampling of certain flow regions.
Additionally, settling alters 
the way in which the particles sample the flow. Specifically, settling decreases spatial clustering for small particle inertia, and increases clustering for large inertia.
\item Caustic singularities result in 
multi-valued particle velocities, causing 
continuum configuration-space descriptions  to fail.  Caustic formation depends 
sensitively on the Stokes number, at small particle inertia, as 
only extreme fluctuations of turbulent strain allow particles to  detach from the flow.
The models show that this sensitivity exists in any flow, but the specifics 
depend on the non-universal tails of the distribution
of fluid-velocity gradients.
The models also explain 
that caustics are responsible for the 
anomalous scaling of the particle-velocity structure functions,
describing large collision velocities.
\item Statistical models predict which properties of the inertial-particle dynamics 
\bmc{depend} on 
the Reynolds number of the turbulent flow and which \bmc{do} not. 
For instance, 
fluctuations of particle separations and relative velocities depend only weakly on the Reynolds 
number. This prediction is borne out by DNS 
and experiments, \bmc{when}     
settling speed and particle inertia are sufficiently small to ignore inertial-range fluctuations. 
\end{enumerate}
\end{summary}

% Future Issues
\begin{issues}[FUTURE QUESTIONS]
\begin{enumerate}
\item Many uncertainties remain regarding the hydrodynamic forces and torques on small particles in turbulence,
concerning the effect of convective fluid inertia, in particular  upon added-mass, history, and lift forces.

\item While the moments of relative velocities 
of heavy particles in turbulence are well understood at small spatial separations, their distribution is not.
To make progress, it is necessary to formulate and analyse non-Gaussian statistical models.

\item The relative dynamics of strongly inertial particles explores the inertial range of turbulence.
This can give rise to intricate dependencies of preferential sampling and relative particle velocities on the Reynolds number of the carrier flow, that remain to be understood.
Formulating statistical models that consistently describe how the large-scale fluid motion
affects smaller scales remains a challenge.

\item Natural flows tend to be inhomogeneous and anisotropic, simply because they contain boundaries.  It is an open question how to model the effect of large-scale structures -- such as mean flow gradients -- on the small-scale dynamics.

\item To predict particle collision rates in turbulence, it is necessary to incorporate interactions into statistical models, including not only hydrodynamic but also electrostatic interactions.
The challenge is to find reliable models for the collision efficiency 
that account for 
particle and fluid inertia, the breakdown of hydrodynamics below the mean free path, droplet deformation, and van-der-Waals interactions.

\item For non-spherical particles, translation and rotation are coupled. Developing statistical models for their collisions requires extending the position-velocity phase space to orientations and angular velocities, and to model hydrodynamic interactions between non-spherical inertial particles.
\end{enumerate}
\end{issues}

%Disclosure
\section*{DISCLOSURE STATEMENT}
The authors are not aware of any affiliations, memberships, funding, or financial holdings that might be perceived as affecting the objectivity of this review.

% Acknowledgements
\section*{ACKNOWLEDGMENTS}
JB acknowledges support from French Investments for the Future (project UCA$^{\rm JEDI}$ ANR-15-IDEX-01), from PRACE (project PRA031), and GENCI (grants TGCC t2016-2as027 and IDRIS 2019-A0062A10800). KG was supported by a grant from Vetenskapsr\aa{}det  (no.~2018-03974). BM was supported by Vetenskapsr\aa{}det (grant no.~2021-4452), and acknowledges a  Mary Shepard B. Upson Visiting Professorship with the Sibley School of Mechanical and Aerospace Engineering at Cornell.
We thank E. Bodenschatz for his encouragement and support, and we acknowledge the hospitality of the KITP in Santa Barbara (USA) where part of this review was written
during the program \textit{Multiphase Flows in Geophysics and the Environment}. Statistical-model simulations were performed on resources provided by the Swedish National Infrastructure for Computing (SNIC). We thank A. Bragg for calculating the KE-model predictions shown in \textbf{Figures~\ref{fig:clustering}} and~\textbf{\ref{fig:rel_vel}},  A. Bhatnagar for sending us the data shown in \textbf{Figure}~\textbf{\ref{fig:rel_vel}d}, and  E. Bodenschatz for providing \textbf{Figure~\ref{fig:preferential}c}.

%\bibliographystyle{ar-style1}
%\bibliography{references}

\begin{thebibliography}{}
\expandafter\ifx\csname natexlab\endcsname\relax\def\natexlab#1{#1}\fi

\bibitem[Abrahamson(1975)]{abrahamson1975collision}
Abrahamson J. 1975.
Collision rates of small particles in a vigorously turbulent fluid.
\textit{Chem. Eng. Sci.} 30:1371--1379

\bibitem[Afonso(2008)]{martins2008terminal}
Afonso MM. 2008.
The terminal velocity of sedimenting particles in a flowing fluid.
\textit{J. Phys. A} 41:385501

\bibitem[Alipchenkov et~al.(2004)Alipchenkov, Zaichik \&
  Petrov]{alipchenkov2004clustering}
Alipchenkov VM, Zaichik LI, Petrov OF. 2004.
Clustering of charged particles in isotropic turbulence.
\textit{High Temperature} 42:919--927

\bibitem[Aliseda et~al.(2002)Aliseda, Cartellier, Hainaux \&
  Lasheras]{aliseda2002effect}
Aliseda A, Cartellier A, Hainaux F, Lasheras JC. 2002.
Effect of preferential concentration on the settling velocity of heavy
  particles in homogeneous isotropic turbulence.
\textit{J. Fluid Mech.} 468:77--105

\bibitem[Ariki et~al.(2018)Ariki, Yoshida, Matsuda \&
  Yoshimatsu]{ariki2018scale}
Ariki T, Yoshida K, Matsuda K, Yoshimatsu K. 2018.
Scale-similar clustering of heavy particles in the inertial range of
  turbulence.
\textit{Phys. Rev. E} 97:033109

\bibitem[Balachandar \& Eaton(2010)]{balachandar2010turbulent}
Balachandar S, Eaton J. 2010.
Turbulent dispersed multiphase flow.
\textit{Annu. Rev. Fluid Mech.} 42:111--133

\bibitem[Balkovsky et~al.(2001)Balkovsky, Falkovich \&
  Fouxon]{balkovsky2001intermittent}
Balkovsky E, Falkovich G, Fouxon A. 2001.
Intermittent distribution of inertial particles in turbulent flows.
\textit{Phys. Rev. Lett.} 86:2790--2793

\bibitem[B\"atge et~al.(2022)B\"atge, Fouxon \&
  Wilczek]{baetge2023quantitative}
B\"atge T, Fouxon I, Wilczek M. 2022.
Quantitative prediction in turbulence at high reynolds numbers.
\textit{arxiv:2208.05384}

\bibitem[Bec et~al.(2006)Bec, Biferale, Boffetta, Cencini, Musacchio \&
  Toschi]{bec2006lyapunov}
Bec J, Biferale L, Boffetta G, Cencini M, Musacchio S, Toschi F. 2006.
Lyapunov exponents of heavy particles in turbulence.
\textit{Phys. Fluids} 18:091702

\bibitem[Bec et~al.(2007)Bec, Biferale, Cencini, Lanotte, Musacchio \&
  Toschi]{bec2007heavy}
Bec J, Biferale L, Cencini M, Lanotte A, Musacchio S, Toschi F. 2007.
Heavy particle concentration in turbulence at dissipative and inertial scales.
\textit{Phys. Rev. Lett.} 98:084502

\bibitem[Bec et~al.(2010)Bec, Biferale, Cencini, Lanotte \&
  Toschi]{bec2010intermittency}
Bec J, Biferale L, Cencini M, Lanotte A, Toschi F. 2010.
Intermittency in the velocity distribution of heavy particles in turbulence.
\textit{J. Fluid Mech.} 646:527--536

\bibitem[Bec et~al.(2005)Bec, Celani, Cencini \& Musacchio]{bec2005clustering}
Bec J, Celani A, Cencini M, Musacchio S. 2005.
Clustering and collisions of heavy particles in random smooth flows.
\textit{Phys. Fluids} 17:073301

\bibitem[Bec et~al.(2008)Bec, Cencini, Hillerbrand \&
  Turitsyn]{bec2008stochastic}
Bec J, Cencini M, Hillerbrand R, Turitsyn K. 2008.
Stochastic suspensions of heavy particles.
\textit{Physica D} 237:2037--2050

\bibitem[Bec \& Ch{\'e}trite(2007)]{bec2007toward}
Bec J, Ch{\'e}trite R. 2007.
Toward a phenomenological approach to the clustering of heavy particles in
  turbulent flows.
\textit{New J. Phys.} 9:77

\bibitem[Bec et~al.(2014)Bec, Homann \& Sankar~Ray]{bec2014gravity}
Bec J, Homann H, Sankar~Ray S. 2014.
Gravity-driven enhancement of heavy particle clustering in turbulent flow.
\textit{Phys. Rev. Lett.} 112:184501

\bibitem[Berry \& Upstill(1980)]{berry1980catastrophe}
Berry M, Upstill C. 1980.
{IV C}atastrophe optics: Morphologies of caustics and their diffraction
  patterns. In \textit{Progress in optics}, ed. E~Wolf, vol.~18. Elsevier,
  257--346

\bibitem[Bewley et~al.(2013)Bewley, Saw \& Bodenschatz]{bewley2013observation}
Bewley GP, Saw EW, Bodenschatz E. 2013.
Observation of the sling effect.
\textit{New J. Phys.} 15:083051

\bibitem[Bhatnagar(2020)]{bhatnagar2020statistics}
Bhatnagar A. 2020.
Statistics of relative velocity for particles settling under gravity in a
  turbulent flow.
\textit{Phys. Rev. E} 101:033102

\bibitem[Bhatnagar et~al.(2018)Bhatnagar, Gustavsson \&
  Mitra]{bhatnagar2018statistics}
Bhatnagar A, Gustavsson K, Mitra D. 2018.
Statistics of the relative velocity of particles in turbulent flows:
  Monodisperse particles.
\textit{Phys. Rev. E} 97:023105

\bibitem[Bhatnagar et~al.(2022)Bhatnagar, Pandey, Perlekar \&
  Mitra]{bhatnagar2022rate}
Bhatnagar A, Pandey V, Perlekar P, Mitra D. 2022.
Rate of formation of caustics in heavy particles advected by turbulence.
\textit{Philos. Trans. Royal Soc. A} 380:20210086

\bibitem[Birnstiel et~al.(2016)Birnstiel, Fang \& Johansen]{birnstiel2016dust}
Birnstiel T, Fang M, Johansen A. 2016.
Dust evolution and the formation of planetesimals.
\textit{Space Sci. Rev.} 205:41--75

\bibitem[Bodenschatz et~al.(2010)Bodenschatz, Malinowski, Shaw \&
  Stratman]{bodenschatz2010can}
Bodenschatz E, Malinowski S, Shaw R, Stratman F. 2010.
Can we understand clouds without turbulence?
\textit{Science} 327:970--971

\bibitem[Boffetta et~al.(2004)Boffetta, De~Lillo \& Gamba]{boffetta2004large}
Boffetta G, De~Lillo F, Gamba A. 2004.
Large scale inhomogeneity of inertial particles in turbulent flows.
\textit{Phys. Fluids} 16:L20--L23

\bibitem[Bragg \& Collins(2014{\natexlab{a}})]{bragg2014new_a}
Bragg AD, Collins LR. 2014{\natexlab{a}}.
New insights from comparing statistical theories for inertial particles in
  turbulence: {I}. {S}patial distribution of particles.
\textit{New J. Phys.} 16:055013

\bibitem[Bragg \& Collins(2014{\natexlab{b}})]{bragg2014new_b}
Bragg AD, Collins LR. 2014{\natexlab{b}}.
New insights from comparing statistical theories for inertial particles in
  turbulence: {II}. {R}elative velocities.
\textit{New J. Phys.} 16:055014

\bibitem[Bragg et~al.(2022)Bragg, Hammond, Dhariwal \&
  Meng]{bragg2021hydrodynamic}
Bragg AD, Hammond AL, Dhariwal R, Meng H. 2022.
Hydrodynamic interactions and extreme particle clustering in turbulence.
\textit{J. Fluid Mech.} 933:A31

\bibitem[Bragg et~al.(2015{\natexlab{a}})Bragg, Ireland \&
  Collins]{bragg2015mechanisms}
Bragg AD, Ireland PJ, Collins LR. 2015{\natexlab{a}}.
Mechanisms for the clustering of inertial particles in the inertial range of
  isotropic turbulence.
\textit{Phys. Rev. E} 92:023029

\bibitem[Bragg et~al.(2015{\natexlab{b}})Bragg, Ireland \&
  Collins]{bragg2015relationship}
Bragg AD, Ireland PJ, Collins LR. 2015{\natexlab{b}}.
On the relationship between the non-local clustering mechanism and preferential
  concentration.
\textit{J. Fluid Mech.} 780:327--343

\bibitem[Brandt \& Coletti(2022)]{brandt2022particle}
Brandt L, Coletti F. 2022.
Particle-laden turbulence: Progress and perspectives.
\textit{Annu. Rev. Fluid Mech.} 54:159--189

\bibitem[Calzavarini et~al.(2008)Calzavarini, Kerscher, Lohse \&
  Toschi]{calzavarini2008dimensionality}
Calzavarini E, Kerscher M, Lohse D, Toschi F. 2008.
Dimensionality and morphology of particle and bubble clusters in turbulent
  flow.
\textit{J. Fluid Mech.} 607:13--24

\bibitem[Candelier et~al.(2023)Candelier, Mehaddi, Mehlig \&
  Magnaudet]{candelier2023second}
Candelier F, Mehaddi R, Mehlig B, Magnaudet J. 2023.
Second-order inertial forces and torques on a sphere in a viscous steady linear
  flow.
\textit{J. Fluid Mech.} 954:A25

\bibitem[Carballido et~al.(2010)Carballido, Cuzzi \&
  Hogan]{carballido2010relative}
Carballido A, Cuzzi JN, Hogan RC. 2010.
{Relative velocities of solids in a turbulent protoplanetary disc}.
\textit{Mon. Notices Royal Astron. Soc.} 405:2339--2344

\bibitem[Chen et~al.(2006)Chen, Goto \& Vassilicos]{chen2006turbulent}
Chen L, Goto S, Vassilicos JC. 2006.
Turbulent clustering of stagnation points and inertial particles.
\textit{J. Fluid Mech.} 553:143--154

\bibitem[Chun et~al.(2005)Chun, Koch, Rani, Ahluwalia \&
  Collins]{chun2005clustering}
Chun J, Koch DL, Rani SL, Ahluwalia A, Collins LR. 2005.
Clustering of aerosol particles in isotropic turbulence.
\textit{J. Fluid Mech.} 536:219--251

\bibitem[Coleman \& Vassilicos(2009)]{coleman2009unified}
Coleman SW, Vassilicos JC. 2009.
A unified sweep-stick mechanism to explain particle clustering in two- and
  three-dimensional homogeneous, isotropic turbulence.
\textit{Phys. Fluids} 21:113301

\bibitem[Crisanti et~al.(1992)Crisanti, Falcioni, Provenzale, Tanga \&
  Vulpiani]{crisanti1992lagrangian}
Crisanti A, Falcioni M, Provenzale A, Tanga P, Vulpiani A. 1992.
Dynamics of passively advected impurities in simple two-dimensional flow
  models.
\textit{Phys. Fluids} 4:1805--1820

\bibitem[Daitche \& T\'el(2011)]{daitche2011memory}
Daitche A, T\'el T. 2011.
Memory effects are relevant for chaotic advection of inertial particles.
\textit{Phys. Rev. Lett.} 107:244501

\bibitem[Derevyanko et~al.(2007)Derevyanko, Falkovich, Turitsyn \&
  Turitsyn]{derevyanko2007lagrangian}
Derevyanko SA, Falkovich G, Turitsyn K, Turitsyn S. 2007.
Lagrangian and eulerian descriptions of inertial particles in random flows.
\textit{Journal of Turbulence} 8:N16

\bibitem[Devenish et~al.(2012)Devenish, Bartello, Brenguier, Collins, Grabowski
  et~al.]{devenish2012droplet}
Devenish BJ, Bartello P, Brenguier JL, Collins LR, Grabowski WW, et~al. 2012.
{Droplet growth in warm turbulent clouds}.
\textit{Q. J. R. Meteorol. Soc.} 138:1401--1429

\bibitem[Dhanasekaran et~al.(2021)Dhanasekaran, Roy \&
  Koch]{dhanasekaran2021collision}
Dhanasekaran J, Roy A, Koch DL. 2021.
{Collision rate of bidisperse spheres settling in a compressional non-continuum
  gas flow}.
\textit{J. Fluid Mech.} 910:A10

\bibitem[Dimotakis(2005)]{dimotakis2005turbulent}
Dimotakis PE. 2005.
Turbulent {M}ixing.
\textit{Annu. Rev. Fluid Mech.} 37:329--356

\bibitem[Dubey et~al.(2022)Dubey, Gustavsson, Bewley \&
  Mehlig]{dubey2022bifurcations}
Dubey A, Gustavsson K, Bewley GP, Mehlig B. 2022.
Bifurcations in droplet collisions.
\textit{Phys. Rev. Fluids} 7:064401

\bibitem[Ducasse \& Pumir(2009)]{ducasse2009inertial}
Ducasse L, Pumir A. 2009.
Inertial particle collisions in turbulent synthetic flows: quantifying the
  sling effect.
\textit{Phys. Rev. E} 80:066312

\bibitem[Elgobashi(2019)]{elgobashi2019direct}
Elgobashi S. 2019.
Direct numerical simulation of turbulent flows laden with droplets or bubbles.
\textit{Annu. Rev. Fluid Mech.} 51:217--244

\bibitem[Elperin et~al.(2002)Elperin, Kleeorin, Lvov, Rogachevskii \&
  Sokoloff]{elperin2002clustering}
Elperin T, Kleeorin N, Lvov VS, Rogachevskii I, Sokoloff D. 2002.
Clustering instability of the spatial distribution of inertial particles in
  turbulent flows.
\textit{Phys. Rev. E} 66:036302

\bibitem[Falkovich et~al.(2003)Falkovich, Fouxon \&
  Stepanov]{falkovich2003statistics}
Falkovich G, Fouxon A, Stepanov G. 2003.
Statistics of turbulence-induced fluctuations of particle concentration, In
  \textit{Sedimentation and sedimentation transport},\ pp.  155--158, Kluwer
  Academic Publishers

\bibitem[Falkovich et~al.(2002)Falkovich, Fouxon \&
  Stepanov]{falkovich2002acceleration}
Falkovich G, Fouxon A, Stepanov M. 2002.
Acceleration of rain initiation by cloud turbulence.
\textit{Nature} 419:151--154

\bibitem[Falkovich et~al.(2001)Falkovich, Gaw\c{e}dzki \&
  Vergassola]{falkovich2001particles}
Falkovich G, Gaw\c{e}dzki K, Vergassola M. 2001.
Particles and fields in fluid turbulence.
\textit{Rev. Mod. Phys.} 73:913--975

\bibitem[Falkovich et~al.(2007)Falkovich, Musacchio, Piterbarg \&
  Vucelja]{falkovich2007inertial}
Falkovich G, Musacchio S, Piterbarg L, Vucelja M. 2007.
Inertial particles driven by a telegraph noise.
\textit{Phys. Rev. E} 76:026313

\bibitem[Falkovich \& Pumir(2004)]{falkovich2004intermittent}
Falkovich G, Pumir A. 2004.
Intermittent distribution of heavy particles in a turbulent flow.
\textit{Phys. Fluids} 16:L47--L50

\bibitem[Fevrier et~al.(2005)Fevrier, Simonin \&
  Squires]{fevrier2005partitioning}
Fevrier P, Simonin O, Squires KD. 2005.
Partitioning of particle velocities in gas-solid turbulent flows into a
  continuous field and a spatially uncorrelated random distribution:
  theoretical formalism and numerical study.
\textit{J. Fluid Mech.} 533:1--46

\bibitem[Fouxon(2011)]{fouxon2011construction}
Fouxon I. 2011.
Construction and description of the stationary measure of weakly dissipative
  dynamical systems.
\textit{arXiv:1110.1625}

\bibitem[Fox(2012)]{fox2012multiphase}
Fox RO. 2012.
Large-eddy-simulation tools for multiphase flows.
\textit{Annu. Rev. Fluid Mech.} 44:47--76

\bibitem[Frisch(1995)]{frisch1995turbulence}
Frisch U. 1995.
{Turbulence: the legacy of {A.N.} {K}olmogorov}.
Cambridge University Press

\bibitem[Fung et~al.(1992)Fung, Hunt, Malik \& Perkins]{fung1992kinematic}
Fung JCH, Hunt JCR, Malik NA, Perkins RJ. 1992.
Kinematic simulation of homogeneous turbulence by unsteady random fourier
  modes.
\textit{J. Fluid Mech.} 236:281--318

\bibitem[Gibert et~al.(2012)Gibert, Xu \& Bodenschatz]{gibert2012where}
Gibert M, Xu H, Bodenschatz E. 2012.
Where do small weakly inertial particles go in a turbulent flow?
\textit{J. Fluid Mech.} 698:160--167

\bibitem[Good et~al.(2014)Good, Ireland, Bewley, Bodenschatz, Collins \&
  Warhaft]{good2014settling}
Good G, Ireland P, Bewley G, Bodenschatz E, Collins L, Warhaft Z. 2014.
Settling regimes of inertial particles in isotropic turbulence.
\textit{J. Fluid Mech.} 759:R3

\bibitem[Goossens(2019)]{goossens2019review}
Goossens WR. 2019.
Review of the empirical correlations for the drag coefficient of rigid spheres.
\textit{Powder Technol.} 352:350--359

\bibitem[Goto \& Vassilicos(2006)]{goto2006self}
Goto S, Vassilicos J. 2006.
Self-similar clustering of inertial particles and zero-acceleration points in
  fully developed two-dimensional turbulence.
\textit{Phys. Fluids} 18:115103

\bibitem[Grabowski \& Wang(2013)]{grabowski2013growth}
Grabowski WW, Wang LP. 2013.
Growth of cloud droplets in a turbulent environment.
\textit{Annu. Rev. Fluid Mech.} 45:293--324

\bibitem[Grassberger(1983)]{grassberger1983generalized}
Grassberger P. 1983.
Generalized dimensions of strange attractors.
\textit{Phys. Lett. A} 97(6):227--230

\bibitem[Guseva et~al.(2016)Guseva, Daitche, Feudel \&
  T\'el]{guseva2016history}
Guseva K, Daitche A, Feudel U, T\'el T. 2016.
History effects in the sedimentation of light aerosols in turbulence: The case
  of marine snow.
\textit{Phys. Rev. Fluids} 1:074203

\bibitem[Gustavsson \& Mehlig(2014)]{gustavsson2014relative}
Gustavsson K, Mehlig B. 2014.
{Relative velocities of inertial particles in turbulent aerosols}.
\textit{Journal of Turbulence} 15:34--69

\bibitem[Gustavsson \& Mehlig(2016)]{gustavsson2016statistical}
Gustavsson K, Mehlig B. 2016.
Statistical models for spatial patterns of heavy particles in turbulence.
\textit{Adv. Phys.} 65:1

\bibitem[Gustavsson et~al.(2015)Gustavsson, Mehlig \&
  Wilkinson]{gustavsson2015analysis}
Gustavsson K, Mehlig B, Wilkinson M. 2015.
Analysis of the correlation dimension of inertial particles.
\textit{Phys. Fluids} 27:073305

\bibitem[Gustavsson et~al.(2008)Gustavsson, Mehlig, Wilkinson \&
  Uski]{gustavsson2008variable}
Gustavsson K, Mehlig B, Wilkinson M, Uski V. 2008.
Variable-range projection model for turbulence-driven collisions.
\textit{Phys. Rev. Lett.} 101:174503

\bibitem[Gustavsson et~al.(2014)Gustavsson, Vajedi \&
  Mehlig]{gustavsson2014clustering}
Gustavsson K, Vajedi S, Mehlig B. 2014.
Clustering of particles falling in a turbulent flow.
\textit{Phys. Rev. Lett.} 112:214501

\bibitem[Hammond \& Meng(2021)]{hammond2021particle}
Hammond A, Meng H. 2021.
Particle radial distribution function and relative velocity measurement in
  turbulence at small particle-pair separations.
\textit{J. Fluid Mech.} 921:A16

\bibitem[Hentschel \& Procaccia(1983)]{hentschel1983infinite}
Hentschel HGE, Procaccia I. 1983.
The infinite number of generalized dimensions of fractals and strange
  attractors.
\textit{Physica D} 8:435--444

\bibitem[Hunt \& Kaloshin(1997)]{hunt1997how}
Hunt BR, Kaloshin VY. 1997.
How projections affect the dimension spectrum of fractal measures.
\textit{Nonlinearity} 10:1031

\bibitem[Ijzermans et~al.(2010)Ijzermans, Meneguz \&
  Reeks]{ijzermans2010segregation}
Ijzermans RHA, Meneguz E, Reeks MW. 2010.
Segregation of particles in incompressible random flows: singularities,
  intermittency and random uncorrelated motion.
\textit{J. Fluid Mech.} 653:99--135

\bibitem[Ireland et~al.(2016{\natexlab{a}})Ireland, Bragg \&
  Collins]{ireland2016effect_a}
Ireland PJ, Bragg AD, Collins LR. 2016{\natexlab{a}}.
The effect of {R}eynolds number on inertial particle dynamics in isotropic
  turbulence. {Part 1. S}imulations without gravitational effects.
\textit{J. Fluid Mech.} 796:617--658

\bibitem[Ireland et~al.(2016{\natexlab{b}})Ireland, Bragg \&
  Collins]{ireland2016effect_b}
Ireland PJ, Bragg AD, Collins LR. 2016{\natexlab{b}}.
{The effect of {R}eynolds number on inertial particle dynamics in isotropic
  turbulence. Part 2. Simulations with gravitational effects}.
\textit{J. Fluid Mech.} 796:659--711

\bibitem[Kaplan \& Yorke(1979)]{kaplan1979chaotic}
Kaplan JL, Yorke JA. 1979.
{Chaotic behavior of multidimensional difference equations}, In
  \textit{{P}roceedings on {F}unctional {D}ifferential {E}quations and
  {A}pproximation of {F}ixed {P}oints (Bonn, 1978)}, vol. 730 of
  \textit{{L}ecture {N}otes in {M}athematics},\ pp.  204--227, Berlin: Springer

\bibitem[Karpi\'nska et~al.(2019)Karpi\'nska, Bodenschatz, Malinowski, Nowak,
  Risius et~al.]{karpinska2019turbulence}
Karpi\'nska K, Bodenschatz JFE, Malinowski SP, Nowak JL, Risius S, et~al. 2019.
Turbulence-induced cloud voids: observation and interpretation.
\textit{Atmos. Chem. Phys.} 19:4991--5003

\bibitem[Klett \& Davis(1973)]{klett1973theoretical}
Klett JD, Davis M. 1973.
{Theoretical collision efficiencies of cloud droplets at small Reynolds
  numbers}.
\textit{J. Atmos. Sci.} 30:107--117

\bibitem[Kraichnan(1968)]{kraichnan1968small}
Kraichnan R. 1968.
Small-scale structure of a scalar field convected by turbulence.
\textit{Phys. Fluids} 11(5):945--953

\bibitem[Landau \& Lifshitz(1987)]{landau1987hydrodynamics}
Landau LD, Lifshitz EM. 1987.
Fluid mechanics.
Pergamon Press, 2nd ed.

\bibitem[Larsen et~al.(2018)Larsen, Shaw, Kostinski \&
  Glienke]{larsen2018fine-scale}
Larsen ML, Shaw RA, Kostinski AB, Glienke S. 2018.
Fine-scale droplet clustering in atmospheric clouds: 3d radial distribution
  function from airborne digital holography.
\textit{Phys. Rev. Lett.} 121:204501

\bibitem[Ledrappier \& Young(1988)]{ledrappier1988dimension}
Ledrappier F, Young LS. 1988.
Dimension formula for random transformations.
\textit{Commun. Math. Phys.} 117:529--548

\bibitem[Legendre \& Magnaudet(1997)]{legendre1997note}
Legendre D, Magnaudet J. 1997.
A note on the lift force on a spherical bubble or drop in a
  low-{R}eynolds-number shear flow.
\textit{Phys. Fluids} 9:3572--3574

\bibitem[Lillo et~al.(2008)Lillo, Cecconi, Lacorata \&
  Vulpiani]{delillo2008sedimentation}
Lillo FD, Cecconi F, Lacorata G, Vulpiani A. 2008.
Sedimentation speed of inertial particles in laminar and turbulent flows.
\textit{Europhys. Lett.} 84:40005

\bibitem[Lovalenti \& Brady(1993)]{lovalenti1993hydrodynamic}
Lovalenti PM, Brady JF. 1993.
The hydrodynamic force on a rigid particle undergoing arbitrary time-dependent
  motion at small {R}eynolds number.
\textit{J. Fluid Mech.} 256:561--605

\bibitem[Lu et~al.(2010)Lu, Nordsiek, Saw \& Shaw]{lu2010clustering_a}
Lu J, Nordsiek H, Saw E, Shaw RA. 2010.
Clustering of charged inertial particles in turbulence.
\textit{Phys. Rev. Lett.} 104:184505

\bibitem[Lu \& Shaw(2015)]{lu2015charged}
Lu J, Shaw RA. 2015.
Charged particle dynamics in turbulence: Theory and direct numerical
  simulations.
\textit{Phys. Fluids} 27:065111

\bibitem[Magnusson et~al.(2022)Magnusson, Dubey, Kearney, Bewley \&
  Mehlig]{magnusson2021collisions}
Magnusson G, Dubey A, Kearney R, Bewley GP, Mehlig B. 2022.
Collisions of micron-sized, charged water droplets in still air.
\textit{arXiv:2106.11543}

\bibitem[Martin \& Meiburg(1994)]{martin1994accumulation}
Martin JE, Meiburg E. 1994.
The accumulation and dispersion of heavy particles in forced twodimensional
  mixing layers. 1. the fundamental and subharmonic cases.
\textit{Phys. Fluids} 6:1116--1132

\bibitem[Mathai et~al.(2016)Mathai, Calzavarini, Brons, Sun \&
  Lohse]{mathai2016microbubbles}
Mathai V, Calzavarini E, Brons J, Sun C, Lohse D. 2016.
Microbubbles and microparticles are not faithful tracers of turbulent
  acceleration.
\textit{Phys. Rev. Lett.} 117:024501

\bibitem[Mathai et~al.(2020)Mathai, Lohse \& Sun]{mathai2020bubbly}
Mathai V, Lohse D, Sun C. 2020.
Bubbly and buoyant particle--laden turbulent flows.
\textit{Ann. Rev. Condens. Matter Phys.} 11:529--559

\bibitem[Mathai et~al.(2015)Mathai, Prakash, Brons, Sun \&
  Lohse]{mathai2015wake}
Mathai V, Prakash VN, Brons J, Sun C, Lohse D. 2015.
Wake-driven dynamics of finite-sized buoyant spheres in turbulence.
\textit{Phys. Rev. Lett.} 115124501

\bibitem[Maxey(2017)]{maxey2017simulation}
Maxey M. 2017.
Simulation methods for particulate flows and concentrated suspensions.
\textit{Annu. Rev. Fluid Mech.} 49:171--193

\bibitem[Maxey(1987)]{maxey1987gravitational}
Maxey MR. 1987.
The gravitational settling of aerosol particles in homogeneous turbulence and
  random flow fields.
\textit{J. Fluid Mech.} 174:441--465

\bibitem[Maxey \& Corrsin(1986)]{maxey1986gravitational}
Maxey MR, Corrsin S. 1986.
Gravitational settling of aerosol particles in randomly oriented cellular flow
  fields.
\textit{J. Atmos. Sci} 43:1112--1134

\bibitem[Maxey \& Riley(1983)]{maxey1983equation}
Maxey MR, Riley JJ. 1983.
Equation of motion for a small rigid sphere in a nonuniform flow.
\textit{Phys. Fluids} 26:883--889

\bibitem[Mazzitelli \& Lohse(2004)]{mazzitelli2004lagrangian}
Mazzitelli IM, Lohse D. 2004.
Lagrangian statistics for fluid particles and bubbles in turbulence.
\textit{New J. Phys.} 6:203

\bibitem[Meibohm et~al.(2020)Meibohm, Gustavsson, Bec \&
  Mehlig]{meibohm2020fractal}
Meibohm J, Gustavsson K, Bec J, Mehlig B. 2020.
Fractal catastrophes.
\textit{New J. Phys.} 22:013033

\bibitem[Meibohm et~al.(2023{\natexlab{a}})Meibohm, Gustavsson \&
  Mehlig]{meibohm2023caustics}
Meibohm J, Gustavsson K, Mehlig B. 2023{\natexlab{a}}.
Caustics in turbulent aerosols form along the {V}ieillefosse line at weak
  particle inertia.
\textit{Phys. Rev. Fluids} 8:024305

\bibitem[Meibohm et~al.(2021)Meibohm, Pandey, Bhatnagar, Gustavsson, Mitra
  et~al.]{meibohm2021paths}
Meibohm J, Pandey V, Bhatnagar A, Gustavsson K, Mitra D, et~al. 2021.
Paths to caustic formation in turbulent aerosols.
\textit{Phys. Rev. Fluids} 6:L062302

\bibitem[Meibohm et~al.(2017)Meibohm, Pistone, Gustavsson \&
  Mehlig]{meibohm2017relative}
Meibohm J, Pistone L, Gustavsson K, Mehlig B. 2017.
Relative velocities in bidisperse turbulent suspensions.
\textit{Phys. Rev. E} 96:061102

\bibitem[Meibohm et~al.(2023{\natexlab{b}})Meibohm, Sundberg, Mehlig \&
  Gustavsson]{meibohm2023non}
Meibohm J, Sundberg L, Mehlig B, Gustavsson K. 2023{\natexlab{b}}.
Caustic formation in a non-{G}aussian model for turbulent aerosols.
\textit{arxiv:2307.10689}

\bibitem[Minier(2016)]{minier2016statistical}
Minier JP. 2016.
Statistical descriptions of polydisperse turbulent two-phase flows.
\textit{Phys. Rep.} 665:1--122

\bibitem[Mizuno et~al.(1988)Mizuno, Markiewicz \& V{\"o}lk]{mizuno1988grain}
Mizuno H, Markiewicz W, V{\"o}lk H. 1988.
Grain growth in turbulent protoplanetary accretion disks.
\textit{Astron. Astrophys.} 195:183--92

\bibitem[Monchaux et~al.(2010)Monchaux, Bourgoin \&
  Cartellier]{monchaux2010preferential}
Monchaux R, Bourgoin M, Cartellier A. 2010.
Preferential concentration of heavy particles: a {V}orono{\"\i} analysis.
\textit{Phys. Fluids} 22(10):103304

\bibitem[Monchaux et~al.(2012)Monchaux, Bourgoin \&
  Cartellier]{monchaux2012analyzing}
Monchaux R, Bourgoin M, Cartellier A. 2012.
Analyzing preferential concentration and clustering of inertial particles in
  turbulence.
\textit{Int. J. Multiphase Flow} 40:1--18

\bibitem[Nair et~al.(2022)Nair, Devenish \& van Reeuwijk]{nair2022effect}
Nair V, Devenish B, van Reeuwijk M. 2022.
Effect of gravity on particle clustering and collisions in decaying turbulence.
\textit{{P}reprint}

\bibitem[Olivieri et~al.(2014)Olivieri, Picano, Sardina, Iudicone \&
  Brandt]{olivieri2014effect}
Olivieri S, Picano F, Sardina G, Iudicone D, Brandt L. 2014.
The effect of the {B}asset history force on particle clustering in homogeneous
  and isotropic turbulence.
\textit{Phys. Fluids} 26:041704

\bibitem[Paladin \& Vulpiani(1987)]{paladin1987anomalous}
Paladin G, Vulpiani A. 1987.
Anomalous scaling laws in multifractal objects.
\textit{Phys. Rep.} 156:147--225

\bibitem[Pan \& Padoan(2010)]{pan2010relative}
Pan L, Padoan P. 2010.
Relative velocity of inertial particles in turbulent flows.
\textit{J. Fluid Mech.} 661:73--107

\bibitem[Pan \& Padoan(2013)]{pan2013turbulence}
Pan L, Padoan P. 2013.
{Turbulence-induced relative velocity of dust particles. I. Identical
  particles}.
\textit{Astrophys. J.} 776:12

\bibitem[Pan \& Padoan(2014)]{pan2014turbulence_c}
Pan L, Padoan P. 2014.
Turbulence-induced relative velocity of dust particles. {IV. T}he collision
  kernel.
\textit{Astrophys. J.} 797:101

\bibitem[Pan et~al.(2014{\natexlab{a}})Pan, Padoan \&
  Scalo]{pan2014turbulence_a}
Pan L, Padoan P, Scalo J. 2014{\natexlab{a}}.
{Turbulence-induced relative velocity of dust particles. {II. T}he bidisperse
  case}.
\textit{Astrophys. J.} 791(1):48

\bibitem[Pan et~al.(2014{\natexlab{b}})Pan, Padoan \&
  Scalo]{pan2014turbulence_b}
Pan L, Padoan P, Scalo J. 2014{\natexlab{b}}.
{Turbulence-induced relative velocity of dust particles. {III. T}he probability
  distribution}.
\textit{Astrophys. J.} 792:69

\bibitem[Pergolizzi(2012)]{pergolizzi2012etude}
Pergolizzi B. 2012.
Etude de la dynamique de particules interielles dans des ecoulements
  aleatoires.
Ph.D. thesis, Laboratoire J.-L. Lagrange.
Universite de Nice-Sophia Antipolis

\bibitem[Perrin \& Jonker(2014)]{perrin2014preferred}
Perrin VE, Jonker HJJ. 2014.
Preferred location of droplet collisions in turbulent flows.
\textit{Phys. Rev. E} 89:033005

\bibitem[Perrin \& Jonker(2015)]{perrin2015relative}
Perrin VE, Jonker HJJ. 2015.
Relative velocity distribution of inertial particles in turbulence: A numerical
  study.
\textit{Phys. Rev. E} 92:043022

\bibitem[Petersen et~al.(2019)Petersen, Baker \&
  Coletti]{petersen2019experimental}
Petersen AJ, Baker L, Coletti F. 2019.
Experimental study of inertial particles clustering and settling in homogeneous
  turbulence.
\textit{J. Fluid Mech.} 864:925--970

\bibitem[Pinsky \& Khain(1995)]{pinsky1995model}
Pinsky M, Khain A. 1995.
A model of a homogeneous isotropic turbulent flow and its application for the
  simulation of cloud drop tracks.
\textit{Geophysical \& Astrophysical Fluid Dynamics} 81:33--55

\bibitem[Pinsky et~al.(1999)Pinsky, Khain \& Shapiro]{pinsky1999collisions}
Pinsky M, Khain A, Shapiro M. 1999.
{Collisions of small drops in a turbulent flow. Part I: Collision efficiency.
  Problem formulation and preliminary results}.
\textit{JAS} 56:2585

\bibitem[Pinsky et~al.(2007)Pinsky, Khain \& Shapiro]{pinsky2007collisions}
Pinsky M, Khain A, Shapiro M. 2007.
Collisions of cloud droplets in a turbulent flow. {P}art {IV}: {D}roplet
  hydrodynamic interaction.
\textit{J. Atmos. Sci.} 64(7):2462

\bibitem[Pope(1994)]{pope1994lagrangian}
Pope S. 1994.
Lagrangian pdf methods for turbulent flows.
\textit{Annu. Rev. Fluid Mech.} 26:23--63

\bibitem[Prasath et~al.(2019)Prasath, Vasan \&
  Govindarajan]{prasath2019accurate}
Prasath SG, Vasan V, Govindarajan R. 2019.
Accurate solution method for the maxey--riley equation, and the effects of
  basset history.
\textit{J. Fluid Mech.} 868:428--460

\bibitem[Pumir \& Wilkinson(2016)]{pumir2016collision}
Pumir A, Wilkinson M. 2016.
{Collisional Aggregation Due to Turbulence}.
\textit{Annu. Rev. Cond. Mat. Phys.} 7:141--170

\bibitem[Ray \& Collins(2011)]{ray2011preferential}
Ray B, Collins LR. 2011.
Preferential concentration and relative velocity statistics of inertial
  particles in navier--stokes turbulence with and without filtering.
\textit{J. Fluid Mech.} 680:488--510

\bibitem[Reeks(2021)]{reeks2021development}
Reeks MW. 2021.
The development and application of a kinetic theory for modeling dispersed
  particle flows.
\textit{J. Fluids Eng.} 143

\bibitem[Riley \& Patterson(1974)]{riley1974diffusion}
Riley JJ, Patterson GS. 1974.
Diffusion experiments with numerically integrated isotropic turbulence.
\textit{Phys. Fluids} 17:292--297

\bibitem[Rosa et~al.(2016)Rosa, Parishani, Ayala \& Wang]{rosa2016settling}
Rosa B, Parishani H, Ayala O, Wang LP. 2016.
Settling velocity of small inertial particles in homogeneous isotropic
  turbulence from high-resolution dns.
\textit{Int. J. Multiphase Flow} 83:217--231

\bibitem[Saffman \& Turner(1956)]{saffman1956collision}
Saffman PG, Turner JS. 1956.
On the collision of drops in turbulent clouds.
\textit{J. Fluid Mech.} 1:16--30

\bibitem[Salazar \& Collins(2009)]{salazar2009two}
Salazar JP, Collins LR. 2009.
Two-particle dispersion in isotropic turbulent flows.
\textit{Annu. Rev. Fluid Mech.} 41(1):405--432

\bibitem[Saw et~al.(2014)Saw, Bewley, Bodenschatz, Sankar~Ray \&
  Bec]{saw2014extreme}
Saw EW, Bewley GP, Bodenschatz E, Sankar~Ray S, Bec J. 2014.
Extreme fluctuations of the relative velocities between droplets in turbulent
  airflow.
\textit{Phys. Fluids} 26:111702

\bibitem[Saw et~al.(2012)Saw, Shaw, Salazar \& Collins]{saw2012spatial_b}
Saw EW, Shaw RA, Salazar JP, Collins LR. 2012.
Spatial clustering of polydisperse inertial particles in turbulence: {II}.
  comparing simulation with experiment.
\textit{New J. Phys.} 14:105031

\bibitem[Sawford(2001)]{sawford2001turbulent}
Sawford B. 2001.
Turbulent relative dispersion.
\textit{Annu. Rev. Fluid Mech.} 33(1):289--317

\bibitem[Schneider et~al.(2017)Schneider, Teixeira, Bretherton, Brient, Pressel
  et~al.]{schneider2017climate}
Schneider T, Teixeira J, Bretherton CS, Brient F, Pressel KG, et~al. 2017.
Climate goals and computing the future of clouds.
\textit{Nat. Clim. Chang.} 7(1):3--5

\bibitem[Shaw(2003)]{shaw2003particle}
Shaw RA. 2003.
{Particle-turbulence interactions in atmospheric clouds}.
\textit{Annu. Rev. Fluid Mech.} 35:183--227

\bibitem[Sigurgeirsson \& Stuart(2002)]{sigurgeirsson2002model}
Sigurgeirsson H, Stuart AM. 2002.
A model for preferential concentration.
\textit{Phys. Fluids} 14(12):4352--4361

\bibitem[Simonin et~al.(2006)Simonin, Zaichik, Alipchenkov \&
  Février]{simonin2006connection}
Simonin O, Zaichik LI, Alipchenkov VM, Février P. 2006.
Connection between two statistical approaches for the modelling of particle
  velocity and concentration distributions in turbulent flow: The mesoscopic
  {E}ulerian formalism and the two-point probability density function method.
\textit{Phys. Fluids} 18:125107

\bibitem[Snyder \& Lumley(1971)]{snyder1971some}
Snyder WH, Lumley JL. 1971.
Some measurements of particle velocity autocorrelation functions in a turbulent
  flow.
\textit{J. Fluid Mech.} 48:41--71

\bibitem[Soldati \& Marchioli(2009)]{soldati2009physics}
Soldati A, Marchioli C. 2009.
Physics and modelling of turbulent particle deposition and entrainment: Review
  of a systematic study.
\textit{Int. J. Multiphase Flow} 35:827--839

\bibitem[Sommerer \& Ott(1993)]{sommerer1993particles}
Sommerer J, Ott E. 1993.
Particles floating on a moving fluid: {A} dynamically comprehensible physical
  fractal.
\textit{Science} 259:335--339

\bibitem[Squires \& Eaton(1991)]{squires1991preferential}
Squires KD, Eaton JK. 1991.
Preferential concentration of particles by turbulence.
\textit{Phys. Fluids A} 3:1169--1178

\bibitem[Sumbekova et~al.(2017)Sumbekova, Cartellier, Aliseda \&
  Bourgoin]{sumbekova2017preferential}
Sumbekova S, Cartellier A, Aliseda A, Bourgoin M. 2017.
Preferential concentration of inertial sub-{K}olmogorov particles: the roles of
  mass loading of particles, stokes numbers, and reynolds numbers.
\textit{Phys. Rev. Fluids} 2:024302

\bibitem[Sundaram \& Collins(1997)]{sundaram1997collision}
Sundaram S, Collins LR. 1997.
Collision statistics in an isotropic particle-laden turbulent suspension.
\textit{J. Fluid. Mech.} 335:75--109

\bibitem[Sundararajakumar \& Koch(1996)]{sundararajakumar1996noncontinuum}
Sundararajakumar RR, Koch DL. 1996.
{Non-continuum lubrication flows between particles colliding in a gas}.
\textit{J. Fluid Mech.} 313:283--308

\bibitem[Tenneti \& Subramaniam(2014)]{tenneti2014particle}
Tenneti S, Subramaniam S. 2014.
Particle-resolved direct numerical simulation for gas-solid flow model
  development.
\textit{Annu. Rev. Fluid Mech.} 46:199--230

\bibitem[Toschi \& Bodenschatz(2009)]{toschi2009lagrangian}
Toschi F, Bodenschatz E. 2009.
Lagrangian properties of particles in turbulence.
\textit{Annu. Rev. Fluid Mech.} 41:375--404

\bibitem[V{\"o}lk et~al.(1980)V{\"o}lk, Jones, Morfill \&
  Roeser]{volk1980collisions}
V{\"o}lk H, Jones F, Morfill G, Roeser S. 1980.
Collisions between grains in a turbulent gas.
\textit{Astron. Astrophys.} 85:316--325

\bibitem[Vo{\ss}kuhle et~al.(2011)Vo{\ss}kuhle, Pumir \&
  L\'ev\^eque]{vosskuhle2011estimating}
Vo{\ss}kuhle M, Pumir A, L\'ev\^eque E. 2011.
Estimating the collision rate of inertial particles in a turbulent flow:
  Limitations of the \lq ghost collision' approximation.
\textit{J. Phys. Conf. Ser.} 318:052024

\bibitem[Vo{\ss}kuhle et~al.(2014)Vo{\ss}kuhle, Pumir, L{\'e}v{\^e}que \&
  Wilkinson]{vosskuhle2013prevalence}
Vo{\ss}kuhle M, Pumir A, L{\'e}v{\^e}que E, Wilkinson M. 2014.
Prevalence of the sling effect for enhancing collision rates in turbulent
  suspensions.
\textit{J. Fluid Mech.} 749:841--852

\bibitem[Vo{\ss}kuhle et~al.(2015)Vo{\ss}kuhle, Pumir, L\'ev\^eque \&
  Wilkinson]{vosskuhle2015collision}
Vo{\ss}kuhle M, Pumir A, L\'ev\^eque E, Wilkinson M. 2015.
Collision rate for suspensions at large stokes numbers - comparing
  {N}avier-{S}tokes and synthetic turbulence.
\textit{Journal of Turbululence} 16:15--25

\bibitem[Voth \& Soldati(2017)]{voth2017anisotropic}
Voth GA, Soldati A. 2017.
Anisotropic particles in turbulence.
\textit{Annu. Rev. Fluid Mech.} 49:249--276

\bibitem[Wang \& Maxey(1993)]{wang1993settling}
Wang L, Maxey MR. 1993.
Settling velocity and concentration distribution of heavy particles in
  homogeneous isotropic turbulence.
\textit{J. Fluid Mech.} 256:27--68

\bibitem[Wang et~al.(2008)Wang, Ayala, Rosa \& Grabowski]{wang2008turbulent}
Wang LP, Ayala O, Rosa B, Grabowski WW. 2008.
Turbulent collision efficiency of heavy particles relevant to cloud droplets.
\textit{New J. Phys.} 10:075013

\bibitem[Warhaft(2000)]{warhaft2000passive}
Warhaft Z. 2000.
Passive scalars in turbulent flows.
\textit{Annu. Rev. Fluid Mech.} 32(1):203--240

\bibitem[Wilkinson \& Mehlig(2005)]{wilkinson2005caustics}
Wilkinson M, Mehlig B. 2005.
Caustics in turbulent aerosols.
\textit{Europhys. Lett.} 71:186--192

\bibitem[Wilkinson et~al.(2007)Wilkinson, Mehlig, {\"O}stlund \&
  Duncan]{wilkinson2007unmixing}
Wilkinson M, Mehlig B, {\"O}stlund S, Duncan KP. 2007.
Unmixing in random flows.
\textit{Phys. Fluids} 19:113303

\bibitem[Yavuz et~al.(2018)Yavuz, Kunnen, Van~Heijst \&
  Clercx]{yavuz2018extreme}
Yavuz M, Kunnen R, Van~Heijst G, Clercx H. 2018.
Extreme small-scale clustering of droplets in turbulence driven by hydrodynamic
  interactions.
\textit{Phys. Rev. Lett.} 120:244504

\bibitem[Yoshimoto \& Goto(2007)]{yoshimoto2007self}
Yoshimoto H, Goto S. 2007.
Self-similar clustering of inertial particles in homogeneous turbulence.
\textit{J. Fluid Mech.} 577:275

\bibitem[Zaichik \& Alipchenkov(2003)]{zaichik2003pair}
Zaichik LI, Alipchenkov VM. 2003.
Pair dispersion and preferential concentration of particles in isotropic
  turbulence.
\textit{Phys. Fluids} 15:1776

\bibitem[{Zel'Dovich}(1970)]{zeldovich1970gravitational}
{Zel'Dovich} YB. 1970.
{Gravitational instability: an approximate theory for large density
  perturbations}.
\textit{Astron. Astrophys.} 500:13--18

\end{thebibliography}

\end{document}